\documentclass[10pt,aps,prd,twocolumn,floats,floatfix,showpacs,superscriptaddress,nofootinbib]{revtex4-2}
\usepackage[utf8]{inputenc}               		

\usepackage{graphicx,mathtools,amssymb,amsmath,amsthm,amsfonts,epsfig,epsf,bm}
\usepackage[outdir=./]{epstopdf}
\usepackage[linktocpage]{hyperref}
\hypersetup{hidelinks}
\hypersetup{pdfstartview=}
\hypersetup{
    colorlinks=true,
    linkcolor=blue,
    filecolor=magenta,      
    urlcolor=BrickRed,
    citecolor=BrickRed,
}
\usepackage[usenames]{color}		
\usepackage{tensor}	
\usepackage{csquotes}
\usepackage{mathrsfs}
\usepackage{autobreak}
\usepackage[caption=false]{subfig}

\newcommand{\dd}{\mathrm{d}}
\newcommand*\DAlembert{\mathop{}\!\mathbin\Box}

\usepackage[dvipsnames]{xcolor}

\begin{document}

\title{Neutron star scalarization with Gauss-Bonnet and Ricci scalar couplings}

\author{\textbf{Giulia Ventagli}}
\affiliation{School of Mathematical Sciences, University of Nottingham,
University Park, Nottingham NG7 2RD, United Kingdom}

\author{\textbf{Georgios Antoniou}}
\affiliation{School of Mathematical Sciences, University of Nottingham,
University Park, Nottingham NG7 2RD, United Kingdom}

\author{\textbf{Antoine Lehébel}}
\affiliation{Centro de Astrof\'{\i}sica e Gravita\c c\~ao  - CENTRA,
Departamento de F\'{\i}sica, Instituto Superior T\'ecnico - IST,
Universidade de Lisboa - UL,
Av. Rovisco Pais 1, 1049-001 Lisboa, Portugal}

\author{\textbf{Thomas P.~Sotiriou}}
\affiliation{School of Mathematical Sciences, University of Nottingham,
University Park, Nottingham NG7 2RD, United Kingdom}
\affiliation{School of Physics and Astronomy, University of Nottingham,
University Park, Nottingham NG7 2RD, United Kingdom}


\begin{abstract}
Spontaneous scalarization of neutron stars has been extensively studied in the Damour and Esposito-Far\`ese model, in which a scalar field couples to the Ricci scalar or, equivalently, to the trace of the energy-momentum tensor. However, scalarization of both black holes and neutron stars may also be triggered by a coupling of the scalar field to the Gauss-Bonnet invariant. The case of the Gauss-Bonnet coupling has also received a lot of attention lately, but the synergy of the Ricci and Gauss-Bonnet couplings has been overlooked for neutron stars. Here, we show that combining both couplings has interesting effects on the properties of scalarized neutron stars, such as affecting their domain of existence or the amount of scalar charge they carry.
\end{abstract}

\maketitle

\section{INTRODUCTION}
Even though General Relativity (GR) is extensively tested in the weak-field regime, it is only recently that we have started constraining it in the strong field regime. Gravitational-wave observations \cite{Abbott:2016blz,TheLIGOScientific:2017qsa,Abbott:2020niy} will soon be rising to the hundreds, providing us with enough data to accurately confront many of the proposed strong gravity GR deviations. Increased precision in observations will allow us to determine whether compact objects, which are associated with extremely large curvatures, have different properties than predicted by GR. 

The phenomenon of \textit{spontaneous scalarization} provides perhaps the most promising framework, in which we can investigate the manifestation of a strong gravity process that remains dormant in low curvature regimes. Spontaneous scalarization was initially proposed in the case of neutron stars by Damour and Esposito-Far\`ese (DEF) \cite{Damour:1992kf,Damour:1993hw}. According to it, a scalar field coupled to gravity in a suitable manner, might acquire a non-trivial structure only in the strong field regime of neutron stars, while remaining trivial and undetected in the weak field regime. In the DEF model black holes do not exhibit scalarization unless it is induced by matter in their vicinity \cite{Hawking:1972qk,Sotiriou:2011dz,Cardoso:2013opa,Cardoso:2013fwa,Palenzuela:2013hsa}. However, recently, a different class of models in which there is scalarization of both black holes and neutron stars has been receiving a lot of attention: scalar-Gauss-Bonnet theories (\textit{e.g.}  \cite{Silva:2017uqg,Doneva:2017bvd,Doneva:2017duq}).

Scalarization of both black holes and neutron stars has been scrutinized in various works concerning many different modifications (bare mass, self-interactions, different field content, \textit{etc} \cite{Ramazanoglu:2016kul,Blazquez-Salcedo:2018jnn,Macedo:2019sem,Herdeiro:2018wub,Ramazanoglu:2017xbl,Ramazanoglu:2018hwk}). Scalarization can be thought of as triggered by a curvature-induced tachyonic instability of the scalar field. In more recent works, it has been shown that this instability can be triggered by spin \cite{Dima:2020yac} and lead to black holes that are scalarized only when rapidly rotating \cite{Herdeiro:2020wei,Berti:2020kgk}. It should be noted that scalarization models differ from certain hairy black hole models ({\em e.g.}~\cite{Sotiriou:2013qea,Sotiriou:2014pfa,Antoniou:2017acq,Antoniou:2017hxj}) in that, in the latter all black holes carry a non-trivial scalar configuration, whereas in the former only black holes with certain mass or spin characteristics deviate from the Kerr metric.

The onset of the tachyonic instability that triggers scalarization is controlled by linear terms (although Ref.~\cite{Doneva:2021tvn} also examined what happens if linear terms are absent from the potential) but eventually this instability is quenched by non-linearities, which control the end-state.
In \cite{Andreou:2019ikc}, all terms that can affect the onset of the instability in the framework of Horndeski theory were listed. However, one of these terms, namely the coupling to the Ricci scalar, has not received much attention in many of the aforementioned works. This is mostly due to the fact that,
in the black-hole scenario, the onset of scalarization is only controlled by the Gauss-Bonnet invariant, since the Ricci scalar evaluates to zero for GR black holes. Nonetheless, including the Ricci term does  seem to provide us with several advantages. To begin with, as discussed in \cite{Antoniou:2020nax}, the Ricci term is crucial if one wants to retrieve a late-time attractor to GR in a cosmological scenario. Additionally, it was shown in \cite{Ventagli:2020rnx} that the Ricci term can help in suppressing the scalarization of neutron stars, which would otherwise tend to place significant constraints. Finally, Ref.~\cite{Antoniou:2021zoy} showed that this term has very interesting effects on the properties of scalarized black holes. Even though the Ricci coupling does not affect the onset of black hole scalarization (being zero in a GR black hole background), it affects the properties of the scalarized solutions and, consequently, observables. For certain values of the Ricci coupling ---~which happen to be consistent with the ones associated with a late-time attractor behaviour~--- the presence of this operator is expected to render black holes radially stable, without the need to introduce self-interaction terms.

For the reasons presented above, it is of great interest to examine how the combination of Ricci and Gauss-Bonnet couplings affects neutron star properties. We present the analytic and numerical setup of our study in Sec.~\ref{sec:setup}. The numerical results are presented in Sec.~\ref{sec:results}. In Sec.~\ref{Sec:parameterSpace}, we determine over which region of the parameter space scalarized solutions exist, for three different stellar scenarios. In Sec.~\ref{Sec:betaNeg} and \ref{Sec:betaPos}, we discuss the properties of the scalarized solutions, in particular their scalar charges and masses. Section \ref{Sec:instabilityLines} investigates in more detail the solutions that always exist near the scalarization thresholds, while Sec.~\ref{Sec:EffMass} explains how, already at the level of the GR solution, a given scalar profile may be favored. We conclude with a discussion in Sec.~\ref{sec:discussion}.

\section{SETUP}
\label{sec:setup}

It has been shown in \cite{Andreou:2019ikc} that, in the framework of Horndeski theories, the minimal action containing all the terms that can affect the onset of a tachyonic instability is
\begin{equation}
\begin{split}\label{eq:ActionCaseI}
S&=\int\dd^4x\sqrt{-g}\bigg\{\dfrac{R}{2\kappa}+X+ \gamma\, G^{\mu\nu}\nabla_\mu\phi\,\nabla_\nu\phi
\\
&\quad -\left(m_\phi^2+\dfrac{\beta}{2} R-\alpha\mathscr{G}\right)\dfrac{\phi^2}{2}\bigg\}
 +S_\mathrm{M},
\end{split}
\end{equation}
where $X=-\nabla_\mu\phi\nabla^\mu\phi/2$, $\kappa=8\pi G/c^4$ and $\mathscr{G}$ is the Gauss-Bonnet invariant
\begin{equation}
\mathscr{G}=R^2-4R_{\mu\nu}R^{\mu\nu}+R_{\mu\nu\rho\sigma}R^{\mu\nu\rho\sigma}.
\end{equation}
 $S_\text{M}$ is the matter action, where matter is assumed to couple minimally to the metric; in other words, we are working in the so-called Jordan frame. $m_\phi$ is the bare mass of the scalar field, and $\alpha$, $\beta$ and $\gamma$ parametrize the deviations from GR. Note that $\beta$ is dimensionless, whereas $\gamma$ and $\alpha$ have the dimension of a length squared. $\beta$ is defined such that it matches the notation of the (linearized) DEF model (see~\cite{Andreou:2019ikc} for a detailed discussion on the relation to the original DEF model). For the purpose of this paper we set $\gamma=0$ and $m_\phi=0$. If a bare mass is included it needs to be tuned to rather small values else it can prevent scalarization altogether \cite{Ramazanoglu:2016kul,Ventagli:2020rnx}, while $\gamma$ has a very limited effect on the threshold of tachyonic scalarization \cite{Ventagli:2020rnx}. Note that by setting these two parameters to zero, we retrieve the action studied in \cite{Antoniou:2021zoy} in the context of spontaneously scalarized black holes. The modified Einstein equation is
 \begin{equation}\label{eq:grav_eq}
    G_{\mu\nu}=\kappa T^\phi_{\mu\nu}+\kappa T^\text{M}_{\mu\nu},
 \end{equation}
where 
\begin{equation}
\begin{split}\label{eq:StressEnScal}
T^{\phi}_{\mu\nu} & =\nabla_\mu\nabla_\nu\phi-\frac{1}{2}g_{\mu\nu}\nabla_\lambda\phi\nabla^\lambda\phi\\
& +\frac{1}{2}\beta\left(G_{\mu\nu}-\nabla_\mu\nabla_\nu+g_{\mu\nu}\nabla_\lambda\nabla^\lambda \right)\phi^2\\
& +2\alpha\big[R\big(\nabla_\mu\nabla_\nu-g_{\mu\nu}\nabla_\lambda\nabla^\lambda\big)\phi^2\\
& +2\big(R_{\mu\nu}\nabla_\lambda\nabla^\lambda-2R_{(\mu\lambda}\nabla_{\nu)}\nabla^\lambda\\
& +4g_{\mu\nu}R_{\lambda\sigma}\nabla^\lambda\nabla^\sigma\big)\phi^2-2R_{\mu\lambda\nu\sigma}\nabla^\lambda\nabla^\sigma\phi^2\big]
\end{split}
\end{equation}
comes from the variation of the $\phi$-dependent part of the action with respect to the metric, and $T^\mathrm{M}_{\mu\nu}=-(2/\sqrt{-g})(\delta S_\mathrm{M}/\delta g^{\mu\nu})$ is the matter stress-energy tensor. The scalar field equation reads
\begin{equation}\label{eq:scal_eq}
    \DAlembert \phi =m_\text{eff}^2\phi,
\end{equation}
where the effective scalar mass is given by
\begin{equation}\label{eq:eff_masss}
    m_\text{eff}^2=\frac{\beta}{2}R-\alpha \mathscr{G}.
\end{equation}
A configuration with a sufficiently\footnote{Any negative effective mass squared will cause an instability in Minkowski spacetime, but a curved spacetime is destabilized only if a certain threshold is exceeded.} negative effective mass squared will suffer from a tachyonic instability, triggering spontaneous scalarization. For the purpose of this paper, we restrict our analysis to static and spherically symmetric spacetimes:
\begin{equation}\label{eq:metric}
\text{d}s^2= - e^{\Gamma(r)}\text{d}t^2+e^{\Lambda(r)}\text{d}r^2+r^2 \text{d}\Omega^2,
\end{equation}
and we assume matter to be described by a perfect fluid with $T^\text{M}_{\mu\nu}=(\epsilon+p)u_\mu u_\nu+p\,g_{\mu\nu}$, where $\epsilon$, $p$ and $u_\mu$ are respectively the energy density, the pressure and the 4-velocity of the fluid. The pressure is  directly related to the energy density through the equation of state. The field equations then take the form of coupled ordinary differential equations for $\Gamma$, $\Lambda$, $\epsilon$ and $\phi$, see Appendix. We can solve algebraically the $(rr)$ component of the modified Einstein equation for $e^\Lambda$. The result is
\begin{equation}\label{eq:ExpLambda}
e^\Lambda=\frac{-B+\delta\sqrt{B^2-4\,A\,C}}{4 A},\,\,\delta=\pm 1
\end{equation}
where
\begin{equation}
\begin{split}
& A=1+\kappa\,r^2p-\frac{1}{2}\,\beta\kappa\phi^2,\\
& B=-2+\beta\kappa\,\phi^2-2\,r\Gamma'+r\beta\kappa\,\phi^2\Gamma'+4\,r\beta\kappa\,\phi\phi'\\
&\qquad -8\,\alpha\kappa\,\phi\Gamma'\phi'+r^2\beta\kappa\phi\Gamma'\phi'+\kappa\,r^2\phi'^2,\\
& C=48\,\alpha\kappa\,\phi\,\Gamma'\phi'.
\end{split}
\end{equation}
For the $\delta=-1$ branch of solutions we do not retrieve GR in the limit $\alpha\to 0$ and $\beta \to 0$, henceforth we will assume $\delta=1$. By substituting Eq.~\eqref{eq:ExpLambda} in the remaining differential equations, we can reduce our problem to an integration in three variables: $\Gamma$, $\phi$ and $\epsilon$. 

\subsection{Expansion for $r\to 0$}\label{Sec:Exp0}
Close to the center of the star, we can perform an analytic expansion of the form 
\begin{equation}\label{eq:smallr}
f(r)=\sum_{n=0}^\infty f_n r^n
\end{equation}
for the functions $\Gamma$, $\Lambda$, $\epsilon$, $p$ and $\phi$.
Plugging these expansions in the field equations, we can solve order by order to determine the boundary conditions at the origin. At this point, there are essentially three quantities that one can freely fix: the central density $\epsilon_0$, the value of the scalar field at the center $\phi_0$, and the value of the time component of the metric at the center, determined by $\Gamma_0$. On the other hand, $\Lambda_0$ has to vanish in order to avoid a conical singularity at the center, while $p_0$ is directly related to $\epsilon_0$ by the equation of state. All higher order quantities $\{\Gamma_i, ...,\phi_i\}$, $i\geq1$ can be determined in terms of the three quantities $\{\epsilon_0,\Gamma_0,\phi_0\}$. We will require that spacetime is asymptotically flat, with a trivial scalar field at spatial infinity, which fixes uniquely $\Gamma_0$ and $\phi_0$, or rather restricts $\phi_0$ to a discrete set of values, each corresponding to a different mode; technically, these values are found through a numerical shooting method. Therefore, for given parameters $\alpha$ and $\beta$, a solution is eventually fully determined by the central density $\epsilon_0$. Different choices of $\epsilon_0$ will translate into different masses. 

We must underline the difference with the black hole case, studied in \cite{Antoniou:2021zoy}. For black holes, the equations are scale invariant up to a redefinition of the couplings. Practically, this means that it is enough to explore the full space of parameters $\alpha$ and $\beta$ for a \textit{fixed} mass. One can then deduce all solutions, of arbitrary mass, by an appropriate rescaling. For neutron stars this scaling symmetry is broken by the equation of state that relates $p$ and $\epsilon$. Therefore, one \textit{a priori} has to explore a 3-dimensional space of parameters ($\epsilon_0$, $\alpha$ and $\beta$) in the case of neutron stars. In order to keep this exploration tractable, as it was already done in \cite{Ventagli:2020rnx}, we will focus our study on a selection of central densities and equations of state. We pick these in order to cover very diverse solutions, typically corresponding to the lightest/heaviest observed stars in general relativity. We then explore a wide range of the $(\alpha,\beta)$ parameter space for these fixed densities and equations of state.

To complete this section, let us note that solving order by order the field equations for the higher order coefficients in the expansion \eqref{eq:smallr} does not always yield solutions. All first order coefficients in this expansion have to vanish; one can express $\Gamma_2$, $\epsilon_2$, $p_2$ and $\phi_2$ in terms of $\Lambda_2$; however, $\Lambda_2$ itself is determined by the following equation:
\begin{widetext}
\begin{equation}\label{eq:Lambda2}
\begin{split}
& \Lambda_2^4(512\,\alpha^3\kappa\,\phi_0^2-256\,\alpha^3\beta\kappa^2\phi_0^4)+\Lambda_2^3(512\,p_0\alpha^3\kappa^2\phi_0^2-64\,\alpha^2\beta\kappa\phi_0^2+32\,\alpha^2\beta^2\kappa^2\phi_0^4)+\Lambda_2^2(12\,\alpha\beta^3\kappa^2\phi_0^4-24\,\alpha\beta^2\kappa\phi_0^2\\
& -192\,p_0\alpha^2\beta\kappa^2\phi_0^2)+\Lambda_2\left(2\,\beta-\frac{16}{3}\alpha\epsilon_0\kappa-2\,\beta^2\kappa\,\phi_0^2+3\,\beta^3\kappa\,\phi_0^2+24\,p_0\alpha\beta^2\kappa^2\phi_0^2+\frac{8}{3}\alpha\beta\epsilon_0\kappa^2\phi_0^2+\frac{16}{3}\alpha\beta^2\epsilon_0\kappa^2\phi_0^2\right.\\
&\left.+\frac{1}{2}\beta^3\kappa^2\phi_0^4-\frac{3}{2}\beta^4\kappa^2\phi_0^4\right)-\frac{2}{3}\beta\epsilon_0\kappa+\frac{16}{9}\alpha\epsilon_0^2\kappa^2-p_0\beta^3\kappa^2\phi_0^2+\frac{1}{3}\beta^2\epsilon_0\kappa^2\phi_0^2-\frac{2}{3}\beta^3\epsilon_0\kappa^2\phi_0^2=0.
\end{split}
\end{equation}
\end{widetext}
Equation \eqref{eq:Lambda2} is a fourth order equation in $\Lambda_2$. Such an equation does not necessarily possess real solutions. Therefore, for any choice of parameters $(\alpha,\beta)$ and initial values $(\epsilon_0,\phi_0)$, we need to check that a real solution to Eq.~\eqref{eq:Lambda2} exists. In particular, we need to check this when implementing the shooting method that will allow us to find the values of $\phi_0$ such that the scalar field is trivial at spatial infinity. Such values might actually not exist in the domain where Eq.~\eqref{eq:Lambda2} possesses real solutions.
In practice, we make sure that each choice of parameters that we consider guarantees not only that Eq.~\eqref{eq:Lambda2} has a positive\footnote{An acceptable solution to Eq.~\eqref{eq:Lambda2} must be positive, otherwise $g_{rr}$ diverges at a finite radius, and consequently the pressure and the energy density diverge as well.} real solution, but that such a solution is connected to the GR one. We discard all other parameter combinations that do not respect such criteria.

\subsection{Expansion at spatial infinity}

We now analyze the asymptotic behaviour of the solutions at spatial infinity. This time, we expand the metric and scalar functions in inverse powers of $r$, and solve order by order.
We impose that the asymptotic value of the scalar field vanishes, that is $\phi(r\to\infty)\equiv\phi_\infty=0$, and that $\Gamma(r\to\infty)=0$. The asymptotic solution then reads
\begin{widetext}
\begin{align}
\begin{split}\label{eq:Asymptotic1}
e^{-\Lambda}&= 1-\frac{2M}{r}+\frac{1}{2}\frac{Q^2\kappa}{r^{2}}(1-2\,\beta\kappa)+\frac{1}{2}\frac{MQ^2\kappa}{r^{3}}(1-3\,\beta)+\frac{1}{12}\frac{Q^2\kappa}{r^{4}}\left[ M^2(8-28\,\beta)+Q^2\beta\kappa(1-5\,\beta+12\,\beta^2)\right]\\
& \quad +\frac{1}{48}\frac{MQ^2\kappa}{r^{5}}\left[ 768\,\alpha+8\,M^2(6-23\,\beta)-Q^2\kappa(1-18\,\beta+77\,\beta^2-156\beta^3)\right]+O(r^{-6}),
\end{split}
\\
\begin{split}\label{eq:Asymptotic2}
e^\Gamma&=1-\frac{2M}{r}+\frac{1}{2}\frac{Q^2\beta\kappa}{r^2}+\frac{1}{6}\frac{MQ^2\kappa}{r^3}(1-3\,\beta)+\frac{1}{r^4}\left[ 4\,M^4-\frac{1}{3}M^2Q^2\kappa(1+3\,\beta)+\frac{1}{8}Q^4\beta^2\kappa^2 \right]\\
& \quad -\frac{1}{r^5}\left\{ 8\,M^5-\frac{1}{30}M^3Q^2\kappa(58-75\beta)-\frac{1}{80}M Q^2\kappa\left[ 512\,\alpha-Q^2\kappa(3+10\,\beta-85\,\beta^2+60\,\beta^3) \right] \right\}+O(r^{-6}),
\end{split}
\\
\begin{split}\label{eq:Asymptotic3}
\phi&= \frac{Q}{r}+\frac{MQ}{r^2}+\frac{1}{12}\frac{Q}{r^3}\left[ 16\,M^2-Q^2\kappa(1-2\,\beta+3\,\beta^2) \right]+\frac{1}{r^4}\left[ 2\,M^3Q-\frac{1}{12}MQ^3\kappa(4-9\,\beta+9\,\beta^2) \right]\\
& \quad +\frac{1}{480}\frac{Q}{r^5}\big\{Q^4\kappa^2(9-40\,\beta+86\,\beta^2-144\,\beta^3+117\,\beta^4) -8M^2\left[ 144\,\alpha+Q^2\kappa(58-140\,\beta+105\,\beta^2) \right] \\
& \quad + 1536\,M^4 \big\} +O(r^{-6}).\vphantom{\dfrac{Q}{r}}
\end{split}
\end{align}
\end{widetext}
where $M$ and $Q$ are free. We identify $M$ as the ADM mass and $Q$ as the scalar charge, in the sense that it dictates the fall-off of the scalar field far away. As one can see from Eqs.~\eqref{eq:Asymptotic1}--\eqref{eq:Asymptotic3}, the contribution from the Ricci coupling dominates the asymptotic behaviour of the solutions over the Gauss-Bonnet coupling. Indeed, terms proportional to $\beta$ enter the expansion already at order $r^{-2}$, whereas $\alpha$-dependent terms arise only at order $r^{-5}$. This expansion is in fact entangled with the boundary conditions at the center of the star, as we already mentioned. For fixed parameters $\alpha$ and $\beta$, the freedom in $M$ directly relates to the freedom in the central density $\epsilon_0$. On the other hand, the fact that only discrete values of $\phi_0$ yield a vanishing scalar field at infinity means that the scalar profile is actually fixed once a central density (or a mass) is chosen. Therefore, $Q$ is fixed as a function of $M$, and does not constitute a free charge; this is sometimes referred to as secondary hair.

The scalar charge constitutes probably the most direct channel to test the theory through observations. Indeed, binaries of compact objects endowed with an asymmetric charge will emit dipolar radiation. This enhances the gravitational-wave emission of such systems: in a Post-Newtonian (PN) expansion, dipolar radiation contributes to the energy flux at order -1PN with respect to the usual quadrupolar GR flux. Generically, this dipolar emission is controlled by the sensitivities of the compact objects, defined as\footnote{The factor of $1/\sqrt{4\pi}$ is added to match the standard definition of the sensitivity in the literature, where a different normalization for the scalar field is generally used.}
\begin{equation}
    \alpha_I=\dfrac{1}{\sqrt{4\pi}}\,\dfrac{\partial\text{ln}M_I}{\partial\phi_0},
\end{equation} 
$M_I$ being the mass of the component $I$, and $\phi_0$ the value of the scalar field at infinity. The observation of various binary pulsars, notably the PSR~J1738+0333 system, allows one to set the following constraint:
\begin{equation}
   | \alpha_A-\alpha_B|\lesssim2\times10^{-3},
      \label{eq:DEFbound}
\end{equation}
where $A$ and $B$ label the two components of the system \cite{Shao:2017gwu,Wex:2020ald}. We can then relate the sensitivity to the scalar charge $Q$, using the generic arguments of \cite{Damour:1992we}. We have
\begin{equation}
   Q_I=-\dfrac{1}{4\pi}\,\dfrac{\partial M_I}{\partial\phi_0}.
   \label{eq:DEFcharge}
\end{equation} 
If there is no accidental coincidence in the charge of the two components of the binary, Eqs.~\eqref{eq:DEFbound}-\eqref{eq:DEFcharge} translate as
\begin{equation}
\left|\dfrac{Q}{M}\right|\lesssim6\times10^{-4}
\label{eq:boundQ}
\end{equation}
for the solutions we consider. Only solutions satisfying this bound on the charge to mass ratio are relevant. It is however a non-trivial task to map this bound onto the parameters of the Lagrangian \eqref{eq:ActionCaseI}. We will do so by exploring the parameter space in Sec.~\ref{sec:results}.

\subsection{Numerical implementation}

We solve the system of three differential equations for the three independent functions $\Gamma$, $\phi$ and $\epsilon$ by starting our integration from $r_0=10^{-5}~\text{km}$. We fix the parameters of the theory $\alpha$ and $\beta$, and the central density $\epsilon_0$, typically to values of order $10^{17}$~kg/m$^3$. Then, we give an initial guess for $\phi_0$, and determine boundary conditions as explained in Sec.~\ref{Sec:Exp0}. The integration will generically give a solution; however, we also demand that the scalar field vanishes at infinity, that is $\phi_\infty=0$. Only a discrete set of $\phi_0$ values will yield $\phi_\infty=0$. Each value corresponds to a different number of nodes of the scalar field in the radial direction. In practice, we integrate up to distances $r_\text{max}=300\, \text{km}$ and we implement a shooting method to select the solutions with $\phi_\infty=0$. Generally, we use Mathematica's built-in function FindRoot.

However, in some cases FindRoot fails to find the right solutions, even if one gives it a limited range $(\phi_{0,\:\text{min}},\phi_{0,\:\text{max}})$ where to look for. When this happens, we resort to  bisection instead. In this latter case, we require that $\phi(r_\text{max})/\phi_0 \leq 10^{-2}$. 

At each stage of the shooting method, we must check that Eq.~\eqref{eq:Lambda2} gives a real positive solution for $\Lambda_2$ that is connected to the GR solution. In some cases, we reach the limit of the region of the parameter space where these criteria are fulfilled before reaching $\phi_\infty=0$. When this is the case, there is no solution associated to the given choice of $\alpha$, $\beta$ and $\epsilon_0$. Note also that, given a set of $\alpha$, $\beta$ and $\epsilon_0$, there is a maximum number of nodes that the solution can have, consequently a maximum number of suitable choices of $\phi_0$ (typically up to three modes in the regions we explore). Solutions with more nodes are encountered only for higher values of the parameters $\alpha$ and $\beta$, or at higher curvatures (that is, at higher $\epsilon_0$).

Given a solution, we extract the value of the ADM mass $M$ and the scalar charge $Q$, as defined in the asymptotic expansion \eqref{eq:Asymptotic1}--\eqref{eq:Asymptotic3}. We then have
\begin{equation}
    \begin{split}
    & M = -\left(\frac{1}{2}r^2\Lambda'\,e^{-\Lambda}\right) \bigg|_{r_\text{max}},\\
    & Q = -\left(r^2 \phi'\right)\big|_{r_\text{max}}.
    \end{split}
\end{equation}

\section{NUMERICAL RESULTS}
\label{sec:results}

\subsection{Existence regions of scalarized solutions}
\label{Sec:parameterSpace}

In this section, we will study the regions where scalarized solutions exist in the $(\alpha,\beta)$ parameter space. We analyze three different neutron star scenarios, which correspond to the three cases studied in \cite{Ventagli:2020rnx}.

\subsubsection{Light star with SLy EOS}
\label{sec:lightSLy}

\begin{figure}[ht]
	\includegraphics[width=1\linewidth]{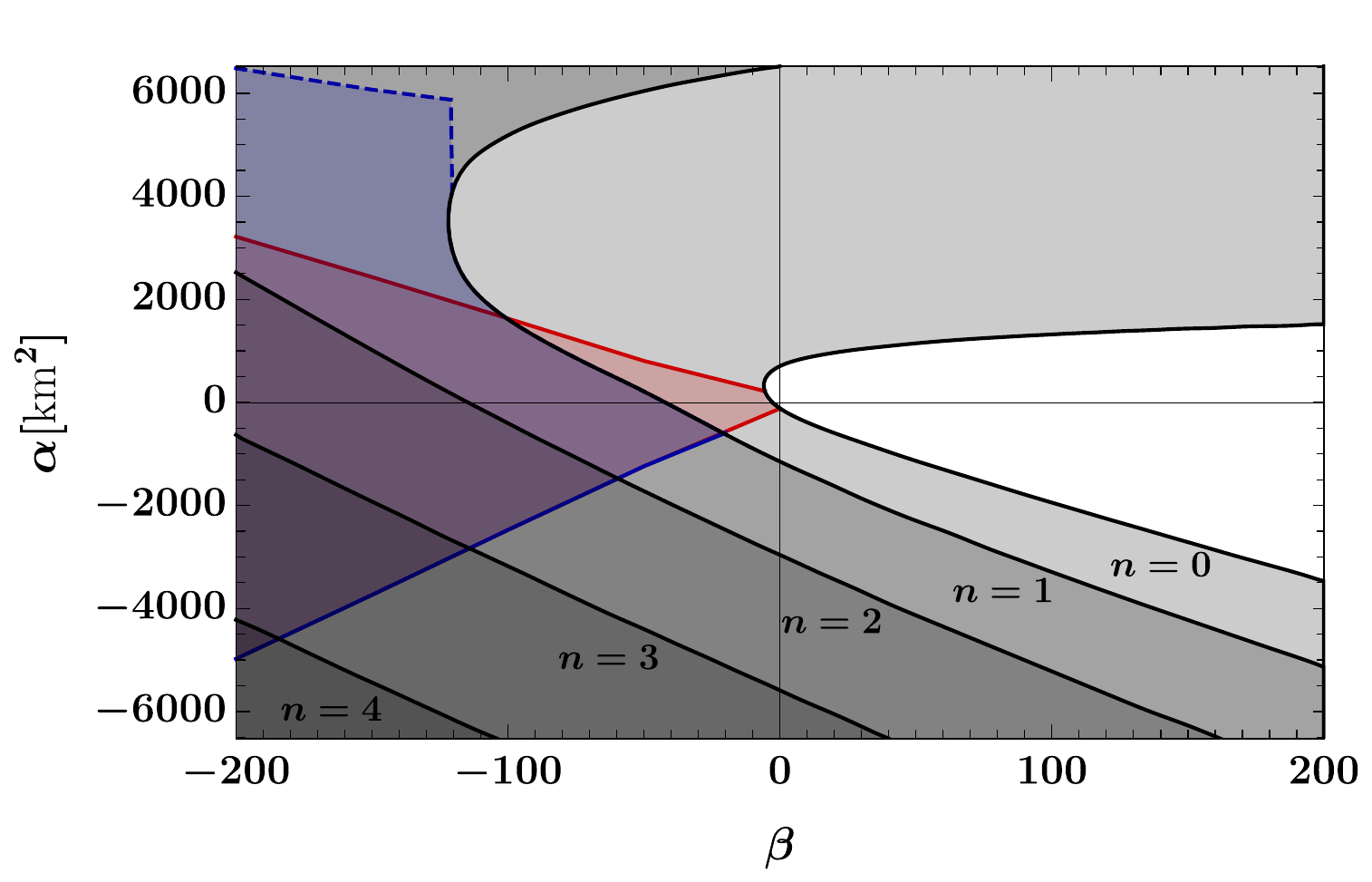}%
	\caption{Regions of existence of scalarized solutions in the $(\alpha,\beta)$ space, for the SLy EOS with $\epsilon_0=8.1\times 10^{17}~\text{kg}/\text{m}^3$. The red (respectively blue) region is the region where scalarized solutions with 0 (respectively 1) node exist. We superimposed the grey contours obtained in Ref.~\cite{Ventagli:2020rnx}, which represent the lines beyond which GR solutions with the same density are unstable to scalar perturbations with 0, 1, 2, \textit{etc} nodes. We see that the region where there exist scalarized solutions with $n$ nodes is included in the region where the GR solutions are unstable to scalar perturbations with $n$ nodes, but much smaller. The dashed boundary for the blue region corresponds to a breakdown of the integration inside the star. In GR, a star with this choice of $\epsilon_0$ and EOS has a light mass, $M_\text{GR}=1.12 M_\odot$.}
	\label{fig:Sly112}
\end{figure}

First, we consider a neutron star described by the SLy equation of state \cite{Haensel:2004nu}, with a central energy density of $\epsilon_0=8.1\times 10^{17}~\text{kg}/\text{m}^3$, so that its gravitational mass in GR is $M_{\text{GR}}=1.12 M_\odot$. The results are summarized in Fig.~\ref{fig:Sly112}, where we relate our new results to the previous study of the scalarization thresholds \cite{Ventagli:2020rnx}.
The white area corresponds to the region of the parameter space where the GR solution is stable. When cranking up the parameters $\alpha$ or $\beta$, a new unstable mode appears every time one crosses a black line. The first mode has 0 nodes, the second 1 node, \textit{etc}. We will refer to these black lines as \textit{instability lines}. Any point in the parameter space that lies within some grey region corresponds to a configuration where the GR solution is unstable. 
The red (respectively blue) area corresponds to the region where scalarized solutions with $n=0$ (respectively $n=1$) nodes exist. We do not include the equivalent regions for higher $n$, to not complicate further the analysis. The region where a scalarized solution does exist
is considerably reduced with respect to the region where the GR solution is unstable.

One of our main results is that the parameters $(\alpha,\beta)$ corresponding to the grey areas that are not covered by the colored regions must be excluded. Indeed, there, scalarized solutions do not exist while the GR solution itself is unstable. Therefore, neutron stars in these theories, when they reach a critical mass, will be affected by a tachyonic instability, but there does not exist a fixed point (a static scalarized solution) where the growth could halt. This would imply that neutron stars with this mass and EOS do not exist for the corresponding parameters  of the theory \eqref{eq:ActionCaseI}. Considering that the properties of the scalarized star are sensitive to nonlinearities, adding further nonlinear interaction terms to the action, \textit{e.g.} self-interactions in a scalar potential, as was proposed in \cite{Macedo:2019sem}, or non-linear terms in the coupling functions \cite{Doneva:2017bvd,Silva:2018qhn}, can potentially change this result.

In Fig.~\ref{fig:Sly112}, the regions where scalarized solutions exist are delimited by \textit{existence lines}, represented by a curve of the respective color. The plain lines correspond to boundaries beyond which it is no longer possible to find a value of $\phi_0$ that allows a suitable solution to Eq.~\eqref{eq:Lambda2}, while providing $\phi_\infty=0$. Beyond dashed lines, on the other hand, nothing special occurs at the center of the star, but the numerical integration breaks down at a finite radius inside the star. We do not know whether, when crossing these dashed lines, our integration is affected by numerical problems, or whether the divergence corresponds to an actual singularity of the solutions. 
It could be that this singularity emerges as an artifact of the method we employ. Indeed, in our analysis, we keep the central density $\epsilon_0$ fixed while pushing the couplings $\alpha$ and $\beta$ to larger and larger values. However, for each couple of parameters $(\alpha,\beta)$, there probably exists a maximal central density beyond which star solutions do not exist, or equivalently it becomes impossible to sustain such a high central density. The dashed line could correspond to this saturation, where we try to push all the parameters beyond values that can actually be sustained by the model.

A surprising feature, which is not visible in Fig.~\ref{fig:Sly112}, is that scalarized solutions always exist in a very narrow range along the instability lines. For example, when crossing the black instability line that delimitates the white region where the GR solution is stable, from the light-grey region where it is unstable against $n=0$ scalar perturbations, there exists a very narrow band (within the grey region) where scalarized solutions with zero node exist. We observed similar behaviours along each instability line, also in the scenarios discussed in the next paragraphs. We further investigate these particular solutions in Sec.~\ref{Sec:instabilityLines}.

\subsubsection{Light star with MPA1 EOS}

\begin{figure}[ht]
	\includegraphics[width=1\linewidth]{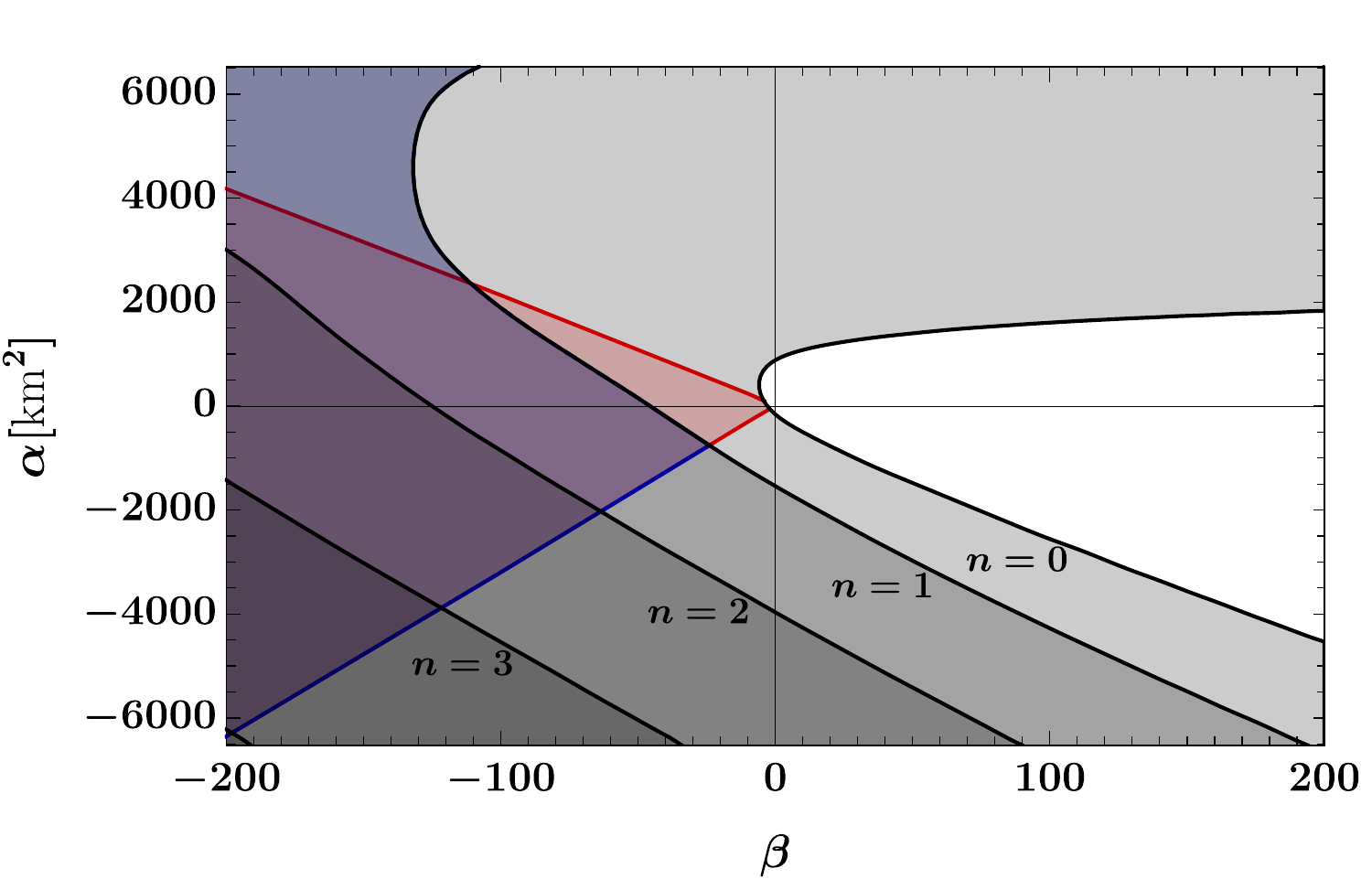}%
	\caption{Regions of existence of scalarized solutions in the $(\alpha,\beta)$ space, for the MPA1 EOS with $\epsilon_0=6.3\times 10^{17}~\text{kg}/\text{m}^3$. The conventions are the same as in Fig.~\ref{fig:Sly112}.  In GR, a star with this choice of $\epsilon_0$ and EOS is again light, with $M_\text{GR}=1.12 M_\odot$.}
	\label{fig:MPA1}
\end{figure}

We next consider a stellar model described by the MPA1 equation of state~\cite{Gungor:2011vq}. We choose a central energy density of $\epsilon_0=6.3\times 10^{17}\,\text{kg}/\text{m}^3$, such that it corresponds to the same GR mass as in the previous case, that is $M_{\text{GR}}=1.12 M_\odot$. We report the results in Fig.~\ref{fig:MPA1}.
As one can see, changing the EOS has only mild effects on the region of existence of scalarized solutions. The analysis of the parameter space is qualitatively the same as for the SLy EOS. The main difference is that, for the range of parameters we considered, no numerical divergences (associated with dashed lines) appear with the MPA1 EOS.

\subsubsection{Heavy star with SLy EOS}

\begin{figure}[ht]
	\subfloat{\includegraphics[width=\linewidth]{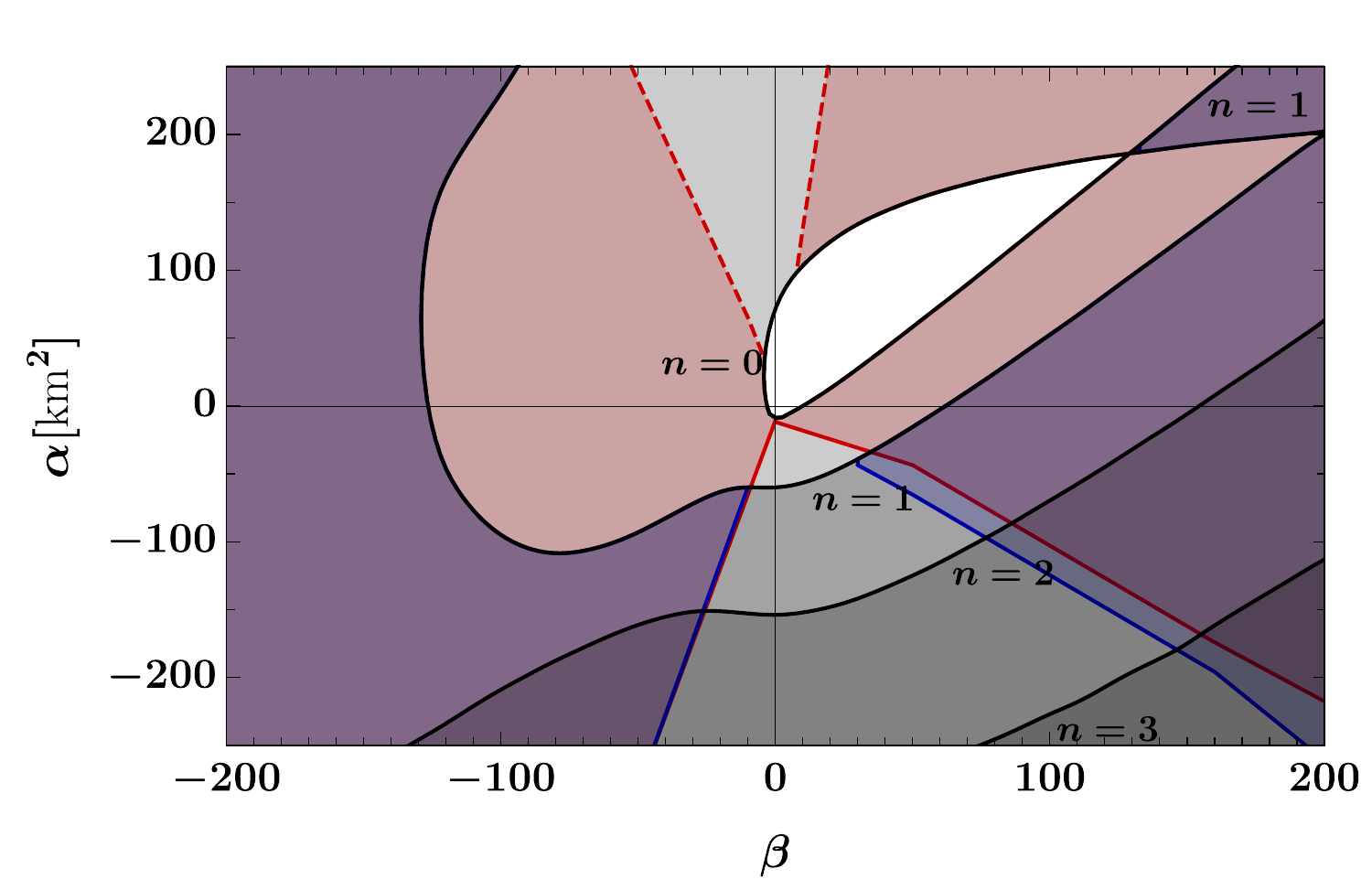}%
	}
	\\
		\subfloat{%
	\includegraphics[width=\linewidth]{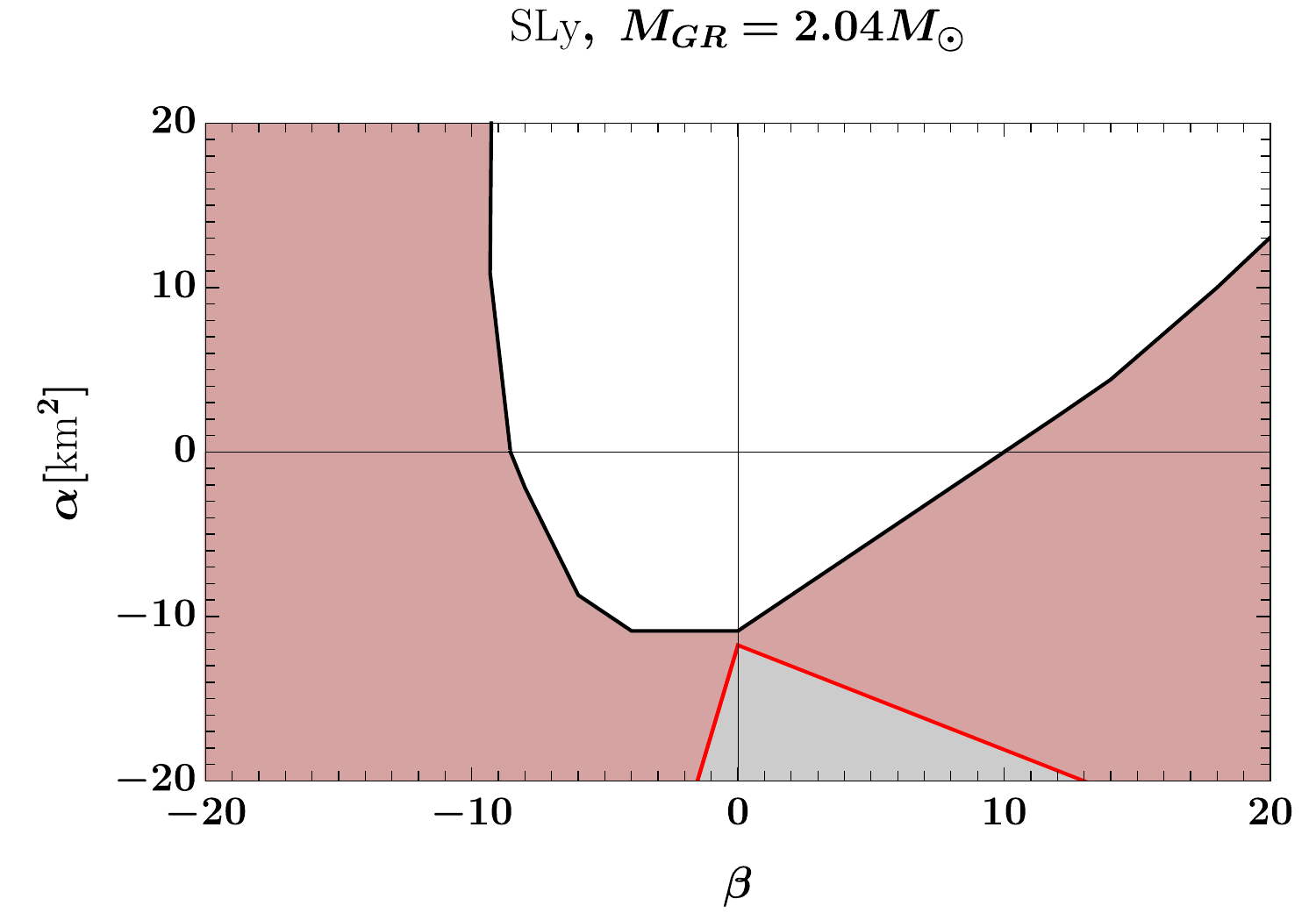}%
	}
	\caption{Regions of existence of scalarized solutions in the $(\alpha,\beta)$ space, for the SLy EOS with $\epsilon_0=3.4\times 10^{18}~\text{kg}/\text{m}^3$. The conventions are the same as in Fig.~\ref{fig:Sly112}. In GR, a star with this choice of $\epsilon_0$ and EOS is the heaviest possible, $M_\text{GR}=2.04 M_\odot$. The bottom panel is simply a zoom of the upper one.}
	\label{fig:SLy204}
\end{figure}

Last, we consider a denser neutron star described by the SLy EOS, with $\epsilon_0=3.4\times 10^{18}\,\text{kg}/\text{m}^3$. It corresponds to an increased mass in GR of $M_{\text{GR}}=2.04 M_\odot$. The results are shown in Fig.~\ref{fig:SLy204}.
In this case, positive values of $\beta$ can also lead to scalarized solutions. Already in \cite{Mendes:2014ufa,Palenzuela:2015ima,Mendes:2016fby,Ventagli:2020rnx}, it was shown that, in GR, dense neutron possess a negative Ricci scalar towards the center, which allows for scalarization to be triggered even when $\beta>0$. As before, a dashed line signals the appearance of divergences, which in this case show up already for the $n=0$ node.

In the lower panel of Fig.~\ref{fig:SLy204}, we zoomed on the region of small couplings, in order to understand better what happens for natural values of the Ricci coupling $\beta$. In the absence of the Gauss-Bonnet coupling, scalarization can occur either if $\beta<-8.55$, or $\beta>11.5$. Let us concentrate on the $\beta>0$ scenario, which is motivated by the results of Ref.~\cite{Antoniou:2020nax}, where it was shown that positive values of $\beta$ make GR a cosmological attractor. We remind that black hole scalarization (at least for non-rotating black holes) occurs for $\alpha>0$. Hence, we see that there exists an interesting region in the $\alpha>0,~\beta>0$ quadrant where even very compact stars do not scalarize, while black holes do. Such models can therefore \textit{a priori} pass all binary pulsar tests, while being testable with black hole observations. On the other hand, for $\beta\gtrsim11.5$, the red region where GR solutions are replaced by scalarized solutions spreads very fast in the $\alpha$ direction, and one has to be careful, when considering black hole scalarization, that such models are not already excluded by neutron star observations.

So far, we established the regions where scalarized solutions exist in the parameter space. In the next two sections, we will discuss the properties of these solutions, in particular their scalar charge and their mass. We separate this study into two cases: $\beta<0$ (Sec.~\ref{Sec:betaNeg}) and $\beta>0$ (Sec.~\ref{Sec:betaPos}); indeed, these two situations have different motivations and observational interests.

\subsection{Mass and scalar charge of the $\beta<0$ solutions}
\label{Sec:betaNeg}

We now focus on the scenario where $\beta<0$. This corresponds to the original situation studied by Damour and Esposito-Farèse. Typically, scalarized solutions with $\beta<0$ and $\alpha=0$ are extremely constrained by binary pulsar observations \cite{Freire:2012mg,Antoniadis:2013pzd,Shao:2017gwu}. A particular motivation to study solutions with $\beta<0$ is therefore to determine whether the addition of a non-zero Gauss-Bonnet coupling can improve their properties. We will consider three different choices of the Ricci coupling: $\beta=-5.5,-10$ and $-100$. The two first choices are relevant astrophysically: $\beta=-5.5$ is approximately the value where scalarization is triggered for small Gauss-Bonnet couplings, while $\beta=-10$ corresponds to a region where neutron stars are scalarized, but with rather small deviations with respect to GR. The third choice, $\beta=-100$, is certainly disfavored observationally, but it will allow us to illustrate an interesting behaviour concerning different scalar modes.

\begin{figure*}[ht]
\begin{center}
	\subfloat{%
	\includegraphics[width=0.4\linewidth]{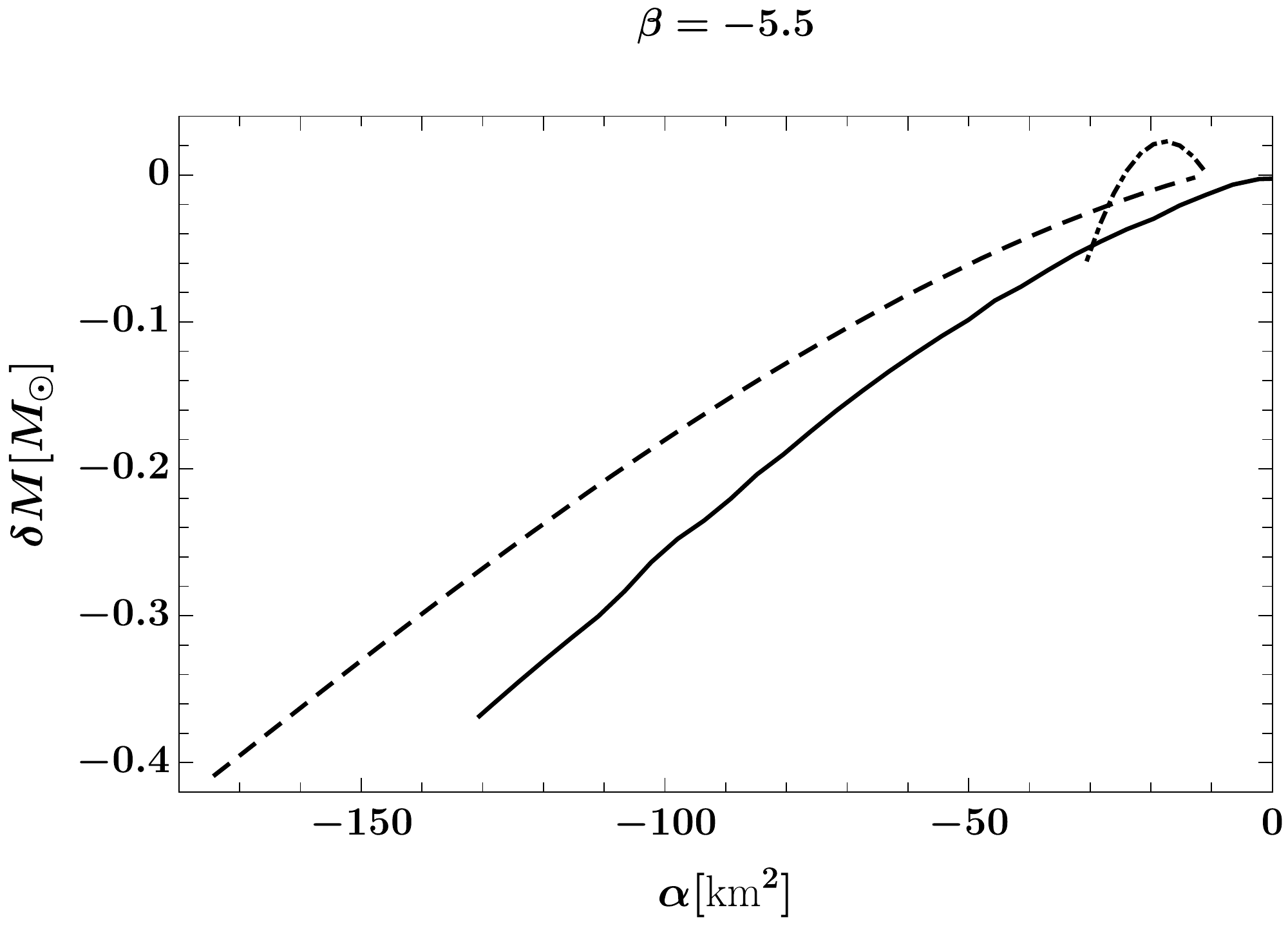}%
	}
	\subfloat{%
	\includegraphics[width=0.4\linewidth]{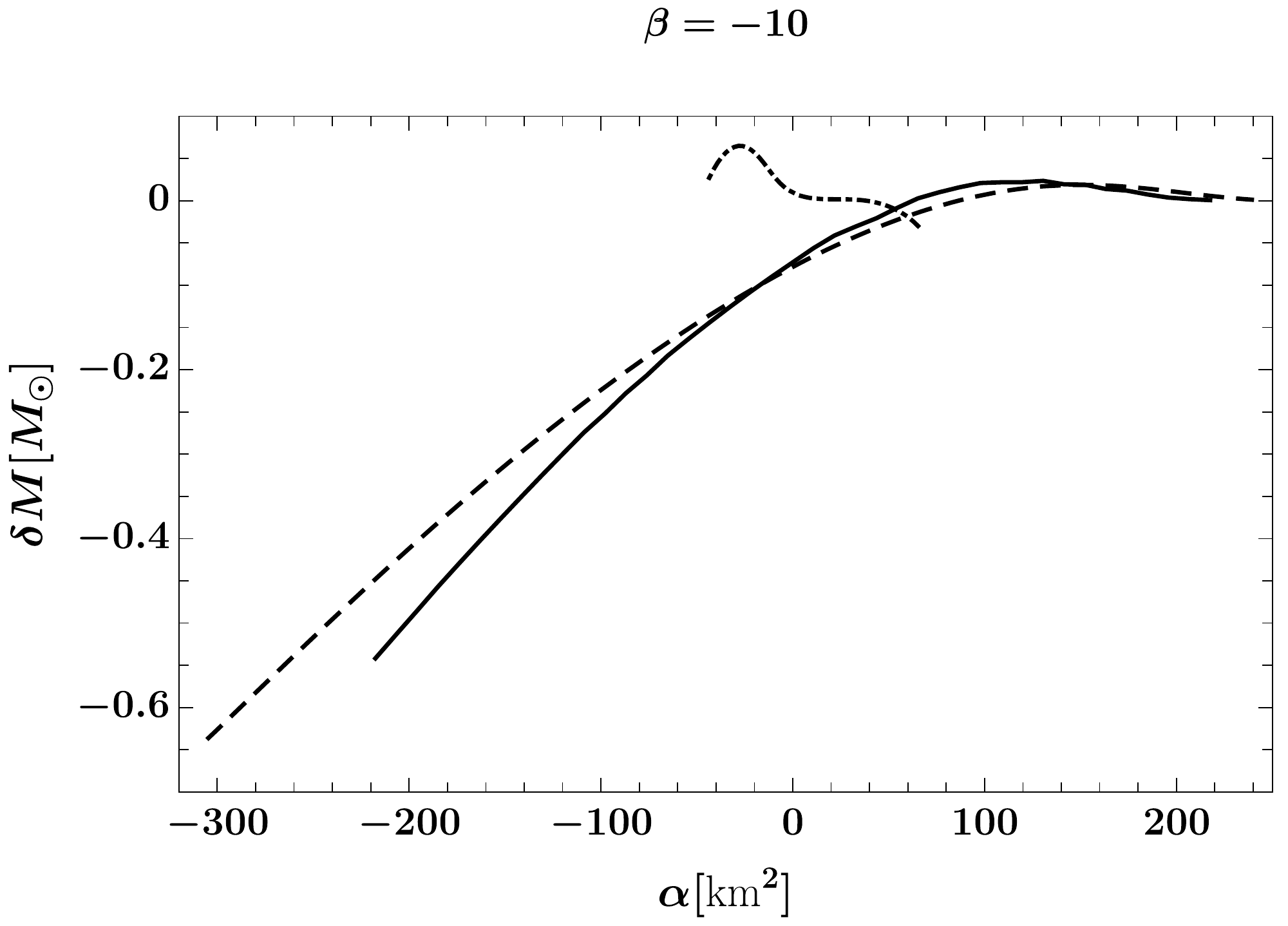}%
	}
	\\
	\subfloat{%
	\includegraphics[width=0.4\linewidth]{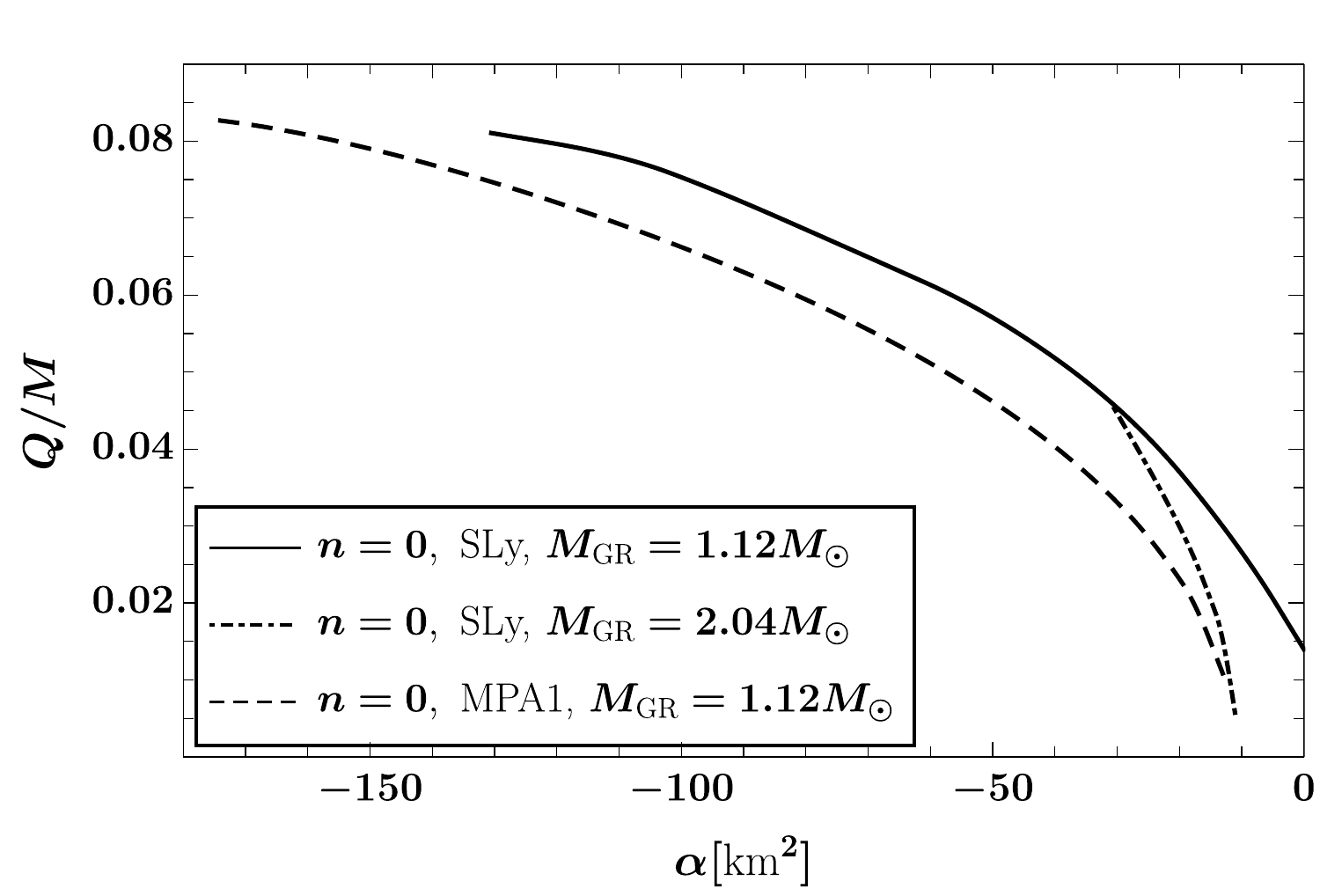}%
	}
	\subfloat{%
	\includegraphics[width=0.4\linewidth]{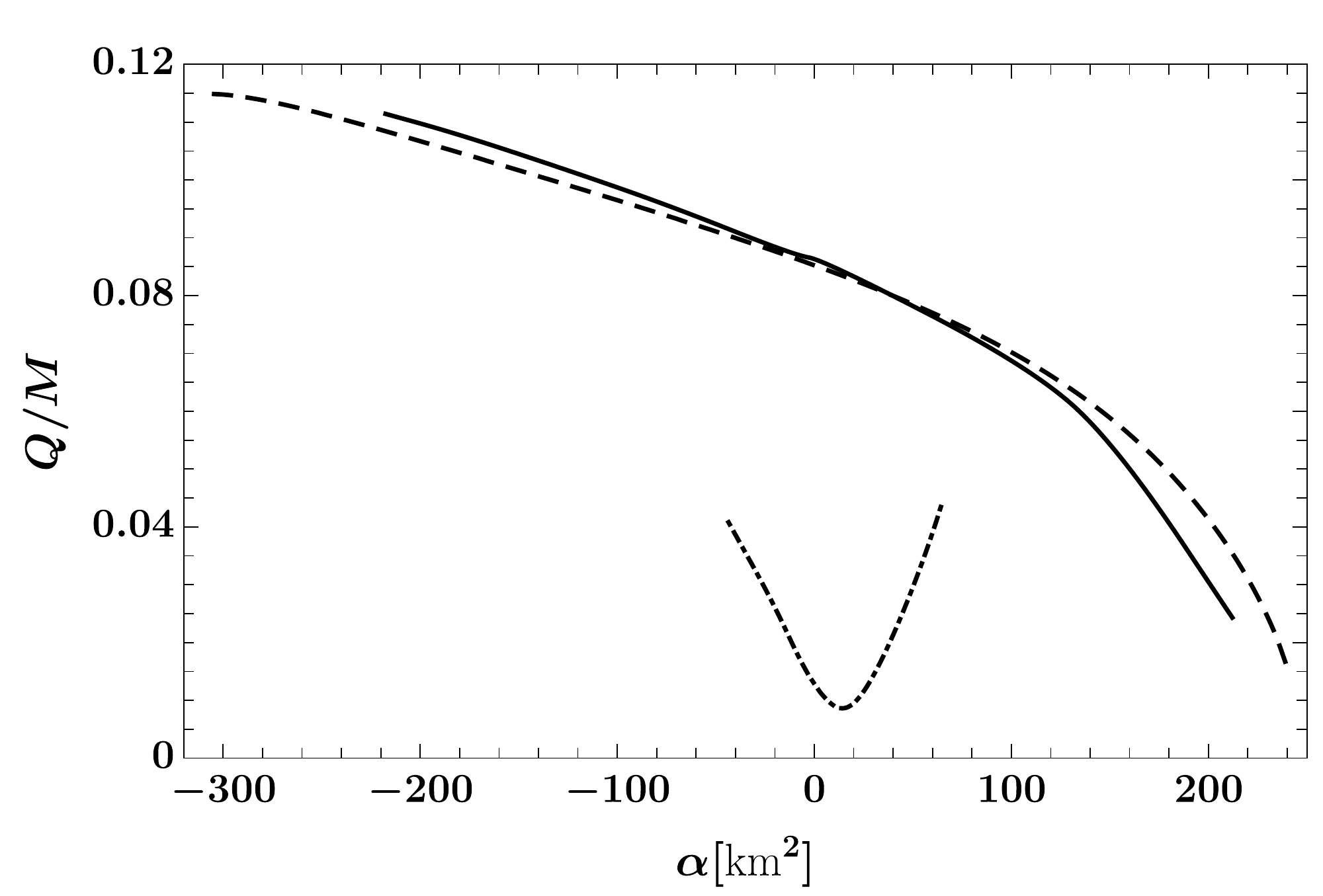}%
	}
	\caption{Mass difference and scalar charge of scalarized solutions for $\beta<0$. The two left (respectively right) panels show how these quantities evolve when varying $\alpha$ at fixed $\beta=-5.5$ (respectively $-10$). The scalar charge $Q$ (bottom panels) is normalized to the total mass of the solutions, $M$. For all curves, the mass difference $\delta M$ (upper panels) is computed with respect to a GR star with the same central density and EOS. Plain curves correspond to a GR mass of $1.12~M_\odot$, using the SLy EOS; dashed curves to the same GR mass, but the MPA1 EOS; and dotted-dashed curves to a GR mass of $2.04~M_\odot$, using the SLy EOS. In this region of the parameter space, only solutions with 0 nodes for the scalar field exist. A generic feature of lighter stars (plain and dashed curves), is that the charge decreases when $\alpha$ increases, \textit{a priori} offering a way to evade the stringent bound of Eq.~\eqref{eq:boundQ} when increasing $\alpha$. However, it is only for values of $\beta$ very close to the DEF threshold ($\beta=-5.5$) that we can obtain scalar charges compatible with observations.
	}
	\label{fig:smallCoup}
\end{center}
\end{figure*}

Let us start with the comparison between the cases $\beta=-5.5$ and $-10$. The results are summarized in Fig.~\ref{fig:smallCoup}.
This figure shows two properties of scalarized stars. First, the mass default (or excess) of scalarized stars with respect to GR stars with the same central density and EOS: 
$\delta M=M-M_{GR}$. 
Second, the scalar charge of the scalarized solutions, $Q$. We compare the results for the three different stellar models considered in Sec.~\ref{Sec:parameterSpace}, for the two values of $\beta$. 
All curves extend only over a finite range of $\alpha$. Indeed, passed a certain value of $\alpha$, we exit the red region on the $\beta<0$ side of Figs.~\ref{fig:Sly112}, \ref{fig:MPA1} and \ref{fig:SLy204} (moving vertically, since $\beta$ is fixed to $-5.5$ or $-10$). Scalarized solutions do not exist outside of this region. 

Figure \ref{fig:smallCoup} shows that the choice of EOS does not affect much the properties of the scalarized solutions.
However, increasing the density drastically modifies these properties. In particular, at higher densities, there exist solutions with $\delta M>0$. This can appear problematic at first. Indeed, one expects that, in a scalarization process, energy is stored in the scalar field distribution. Hence, the ADM mass, that constitutes a measure of the gravitational energy, should decrease in the process. 
However, we stress that we are not studying a dynamical process. Indeed, the stars for which we are computing the mass difference $\delta M$ have, by construction, the same central energy density $\epsilon_0$. In the scalarization process of a GR neutron star, the central energy density will not remain fixed. Hence, our results do not necessarily mean that a star will gain mass when undergoing scalarization.

Perhaps more interestingly for observations, Fig.~\ref{fig:smallCoup} also shows the behaviour of the scalar charge. For the light neutron stars, the scalar charge always decreases when $\alpha$ increases. Therefore, the constraint on the scalar charge, Eq.~\eqref{eq:boundQ}, disfavors the solutions with $\alpha<0$ with respect to standard DEF ($\alpha=0$) solutions. On the contrary, one could hope that a positive Gauss-Bonnet coupling could help evade these constraints even for $\beta<-5.5$, by quenching the charge. Effectively, there will be a direction in the $\alpha>0$ and $\beta<0$ quadrant where the effects of the two operators, Ricci and Gauss-Bonnet, combine to yield a small scalar charge.
This interesting possibility is moderated by what happens in the case of denser stars (dotted-dashed line in Fig.~\ref{fig:smallCoup}). For large negative values of the Ricci coupling ($\beta=-10$), the scalar charge does not have a monotonic behaviour with $\alpha$. In particular, as shown in the bottom-right panel of Fig.~\ref{fig:smallCoup}, $Q$ starts increasing for positive values of $\alpha$. Even at the point where $Q$ is minimal, its value ($Q/M\simeq8\times10^{-3}$) already exceeds the bound of Eq.~\eqref{eq:boundQ}. Therefore, it is only for values of $\beta$ that are very close to the DEF threshold $\beta\simeq-5.5$, that the addition of the Gauss-Bonnet coupling can help to reduce the scalar charge, and to pass the stringent binary pulsar tests.

\begin{figure}[ht]
	\includegraphics[width=0.75\linewidth]{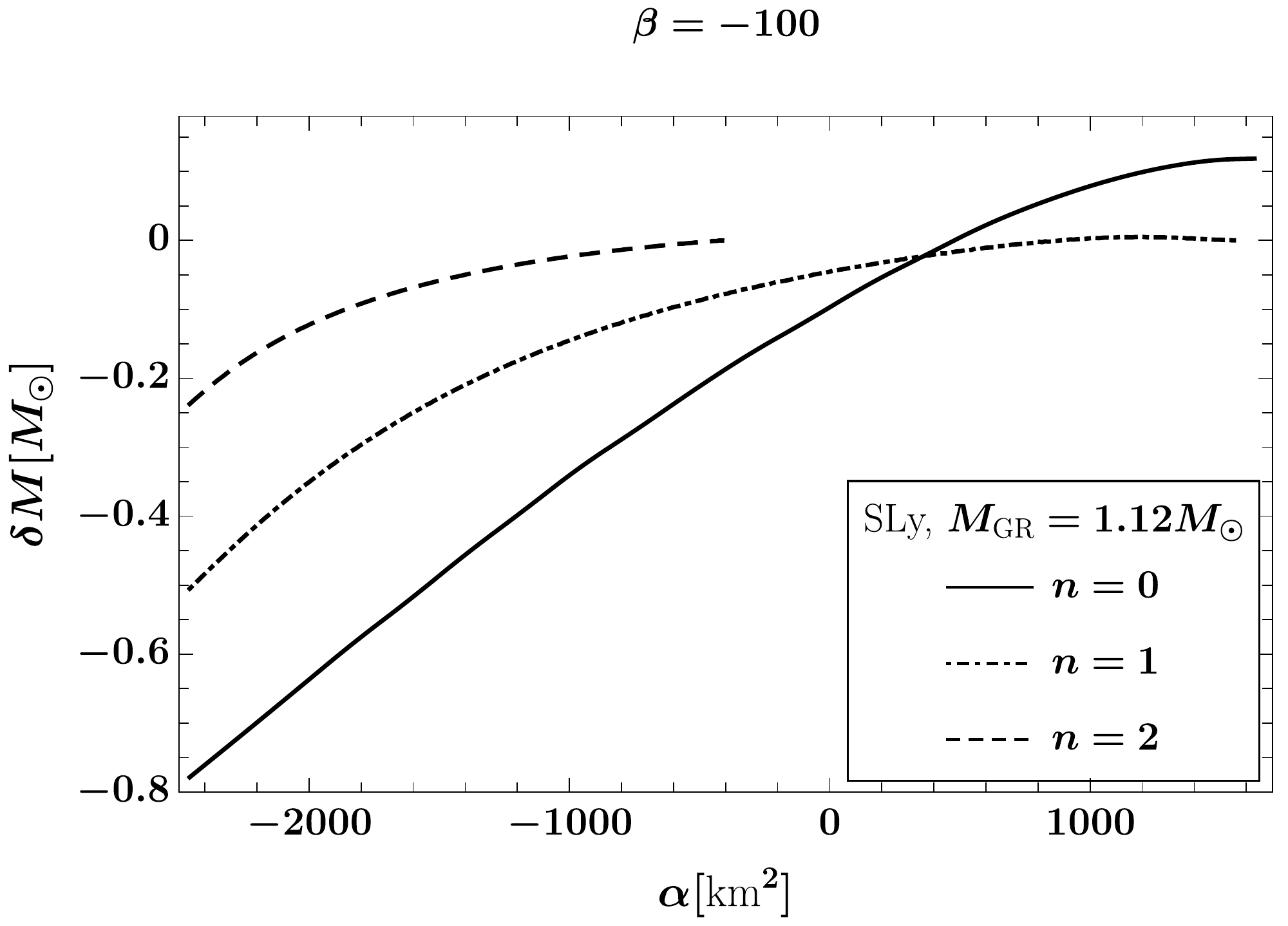}
	\caption{Mass difference $\delta M$ vs $\alpha$ at $\beta=-100$. The EOS considered here is the SLy one, with $\epsilon_0=8.1\times 10^{17}\,\text{kg}/\text{m}^3$, which in GR corresponds to $M_\text{GR}=1.12~M_\odot$. The color and dashing conventions is the same as in Fig.~\ref{fig:smallCoup}. We have more modes in this region of parameter space, that we represent as dotted-dashed (for $n=1$ node) and dashed (for $n=2$ nodes) curves. For $\alpha\gtrsim350\, \text{km}^2$, solutions with 1 node start having a smaller mass than solutions with 0 node, which can indicate that solutions with 1 node are more energetically favored.}
	\label{fig:betaNeg100}
\end{figure}

To conclude the study of the $\beta<0$ region, we consider a significantly more negative Ricci coupling, namely $\beta=-100$. To illustrate what happens at these large negative values of $\beta$, it is enough to consider one scenario, for example the one of lighter neutron stars with the SLy EOS. For such negative values of $\beta$, there exist several scalarized solutions, with different number of nodes. We can then compare the mass difference of these solutions between each other. Figure \ref{fig:betaNeg100} shows that, for $\alpha>\alpha_\text{c}\simeq350\, \text{km}^2$, scalarized solutions with 1 node become lighter than scalarized solutions with 0 node.
This is a hint that, for $\alpha>\alpha_c$, the one node solution will be preferred energetically to the zero node solution. We cannot conclude definitively on this issue, as the ADM mass does not take into account the energy stored in the scalar distribution (which is non-zero for the two scalarized solutions). However, in the regime where this inversion happens, the mass difference with respect to GR, $\delta M$, is rather small. If our interpretation in terms of energetic preference is correct, the transition from a preferred solution with zero node to a solution with one node is interesting. Indeed, the scalarized solution with zero node is associated with the fundamental mode of the GR background instability. At the perturbative level, all the other modes of instability have higher energies. It would then be natural to expect that, at the non-linear level of scalarized solutions, this energy hierarchy is respected. This is the case up to $\alpha=\alpha_\text{c}$, but not anymore beyond. In Sec.~\ref{Sec:EffMass}, we provide a putative explanation for this inversion: that for $\alpha>\alpha_\text{c}$, the profile of the effective mass over the GR background tends to favor the growth of scalar field solutions with one node, rather than zero.

\subsection{Mass and scalar charge of the $\beta>0$ solutions}\label{Sec:betaPos}

\begin{figure}[ht]
	\subfloat{\includegraphics[width=0.75\linewidth]{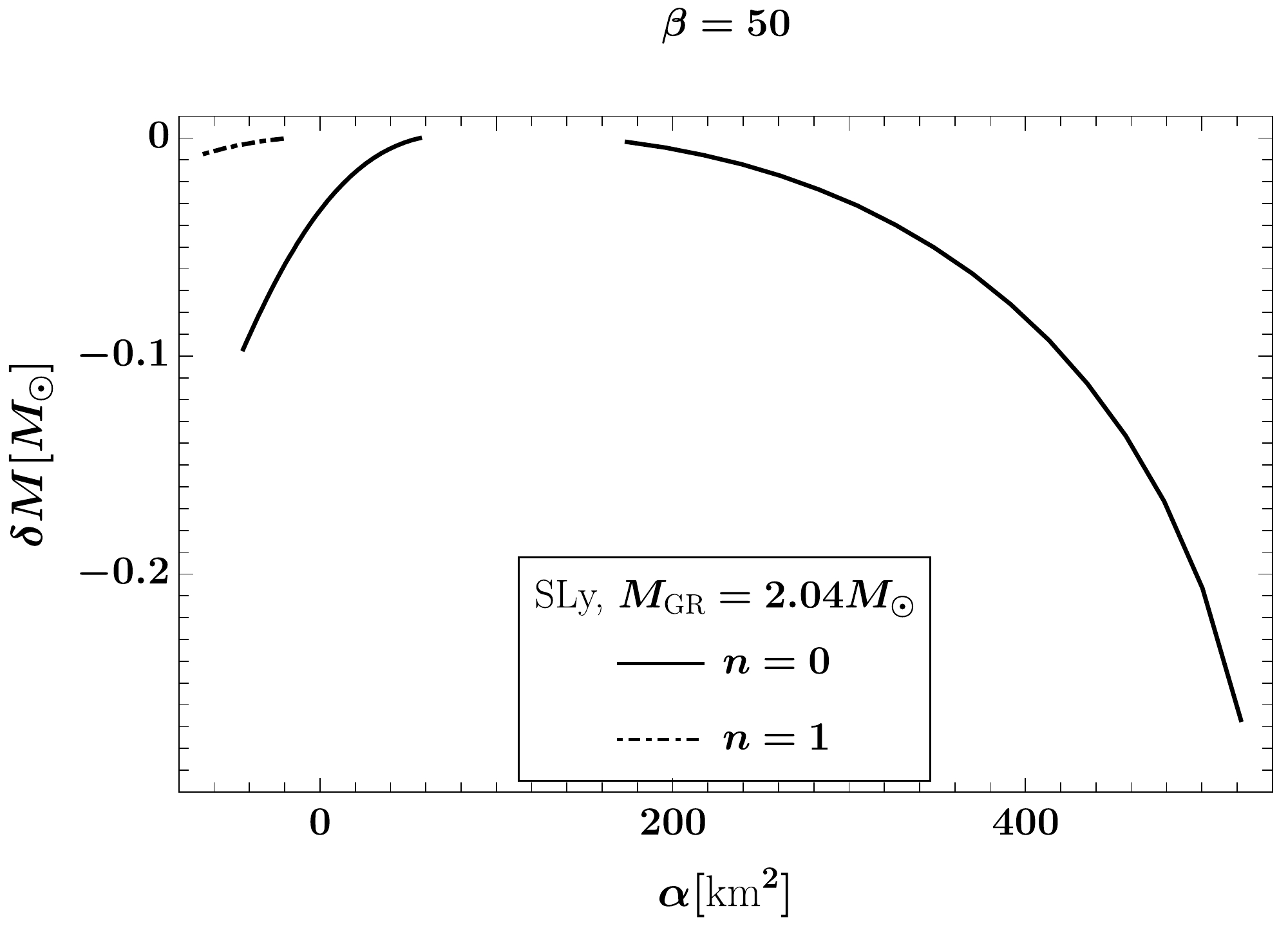}}
	\\
	\subfloat{\includegraphics[width=0.75\linewidth]{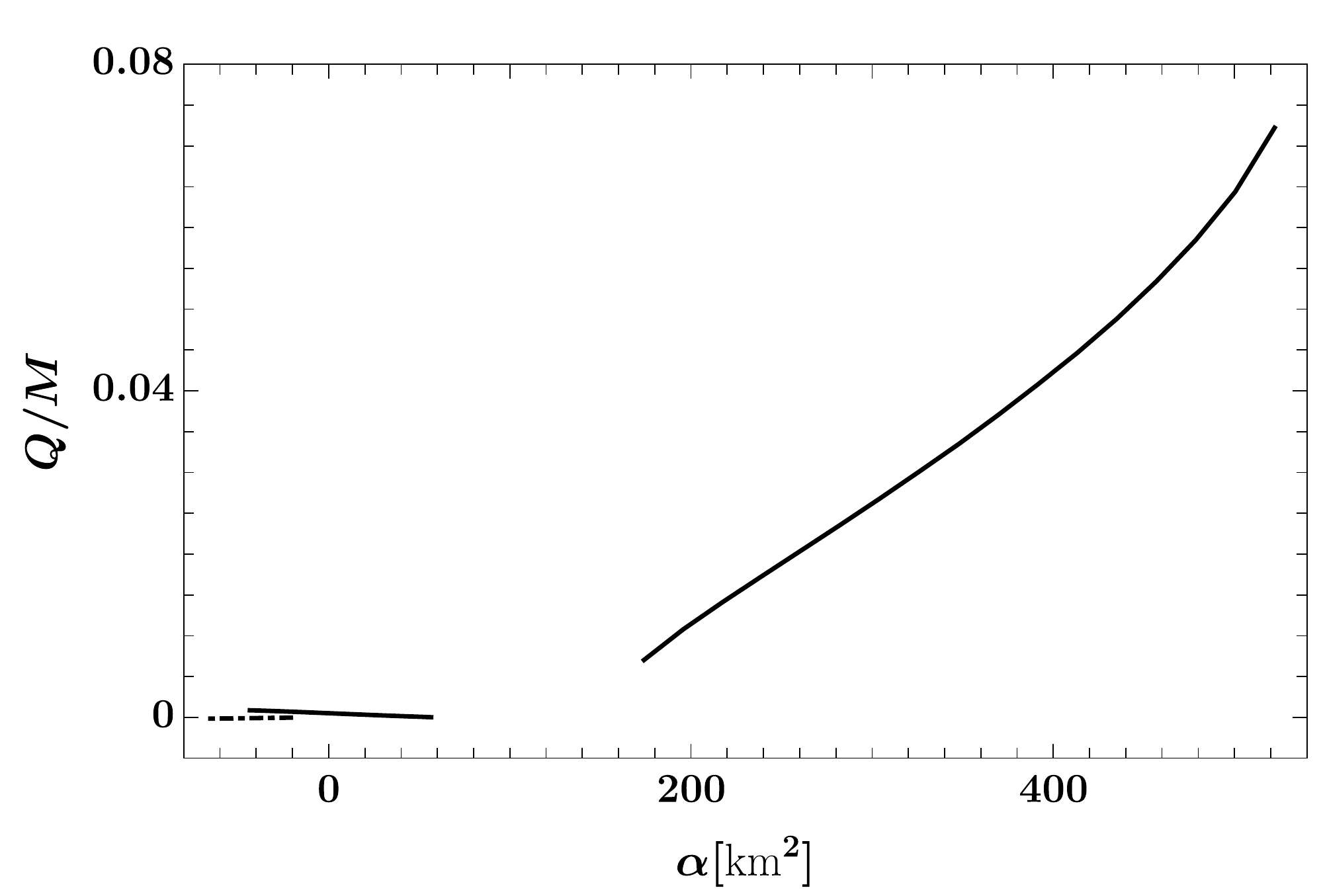}}      
	\caption{Mass difference and scalar charge of scalarized solutions for $\beta>0$ ($\beta=50$ here). Among the three neutron star scenarios that we considered throughout the paper, only the heavier star ($\epsilon_0=5.51 \times 10^{-3}$~kg/m$^3$, $M_\text{GR}=2.04~M_\odot$, SLy EOS) possesses some scalarized solutions in this region. The dashing convention is the same as in Fig.~\ref{fig:betaNeg100}. Solutions that correspond to the interval of $\alpha$ centered on 0 are interesting observationally, as they yield very small scalar charges, compatible with Eq.~\eqref{eq:boundQ}.}
	\label{fig:50M204}
\end{figure}

We now consider the case of positive $\beta$. Such solutions are less constrained by observations than their $\beta<0$ counterparts. They are also very interesting from a cosmological perspective, where $\beta>0$ allows a consistent history throughout different epochs \cite{Antoniou:2020nax}. We have seen in Sec.~\ref{Sec:parameterSpace} that, among the three different possible neutron star configurations we focus on, only the denser one leads to scalarized solutions for $\beta>0$. In Fig.~\ref{fig:50M204}, we show the mass difference $\delta M$ and scalar charge $Q$ as functions of $\alpha$ when $\beta=50$.
Note that scalarized solutions with zero node exist over two disconnected ranges of $\alpha$ ($-44~\text{km}^2<\alpha<57~\text{km}^2$ and $174~\text{km}^2<\alpha<522~\text{km}^2$). In the gap, GR solutions are stable and no scalarized solutions exist. This is obvious from Fig.~\ref{fig:SLy204}, taking a cut along the vertical line $\beta=50$. 
    
Over the first interval, $\alpha$ is rather small and the scalarization process is dominated by the negative Ricci scalar. For strictly vanishing $\alpha$, the scalarization phenomenon with $\beta>0$ has already been examined in \cite{Mendes:2014ufa,Palenzuela:2015ima,Mendes:2016fby}. Here, we find that, in the interval of small values of $\alpha$, the scalar charges of the $n=0$ solutions (as well as of the $n=1$ solutions) are very small. Typically, $Q/M \simeq 10^{-4}-10^{-5}$, compatible with Eq.~\eqref{eq:boundQ}. Hence, all solutions  with $\beta>0$ and rather small values of $\alpha$ are interesting observationally: they display either no scalarization effects for neutron stars (for $\beta\lesssim 11.51$) or very mild scalar charges (for $\beta\gtrsim 11.51$). At the same time, they allow for a consistent cosmological history; finally, together with positive values of $\alpha$, they will generically give rise to black hole scalarization, as studied in detail in \cite{Antoniou:2021zoy}. In this region of parameter space, we can therefore hope to discover scalarization effects in the future gravitational-wave signals of binary black holes, that are either absent or suppressed in the case of neutron stars.
    
Over the second interval ($174~\text{km}^2<\alpha<522~\text{km}^2$), the contribution of the Gauss-Bonnet invariant tends to dominate, and the scalar charges are more significant, as one can immediately notice in Fig.~\ref{fig:SLy204}. Such setups are not compatible with Eq.~\eqref{eq:boundQ}, and therefore less interesting phenomenologically.

\subsection{Scalarized solutions along the instability lines}\label{Sec:instabilityLines}

\begin{figure}[ht]
	\subfloat{\includegraphics[width=0.75\linewidth]{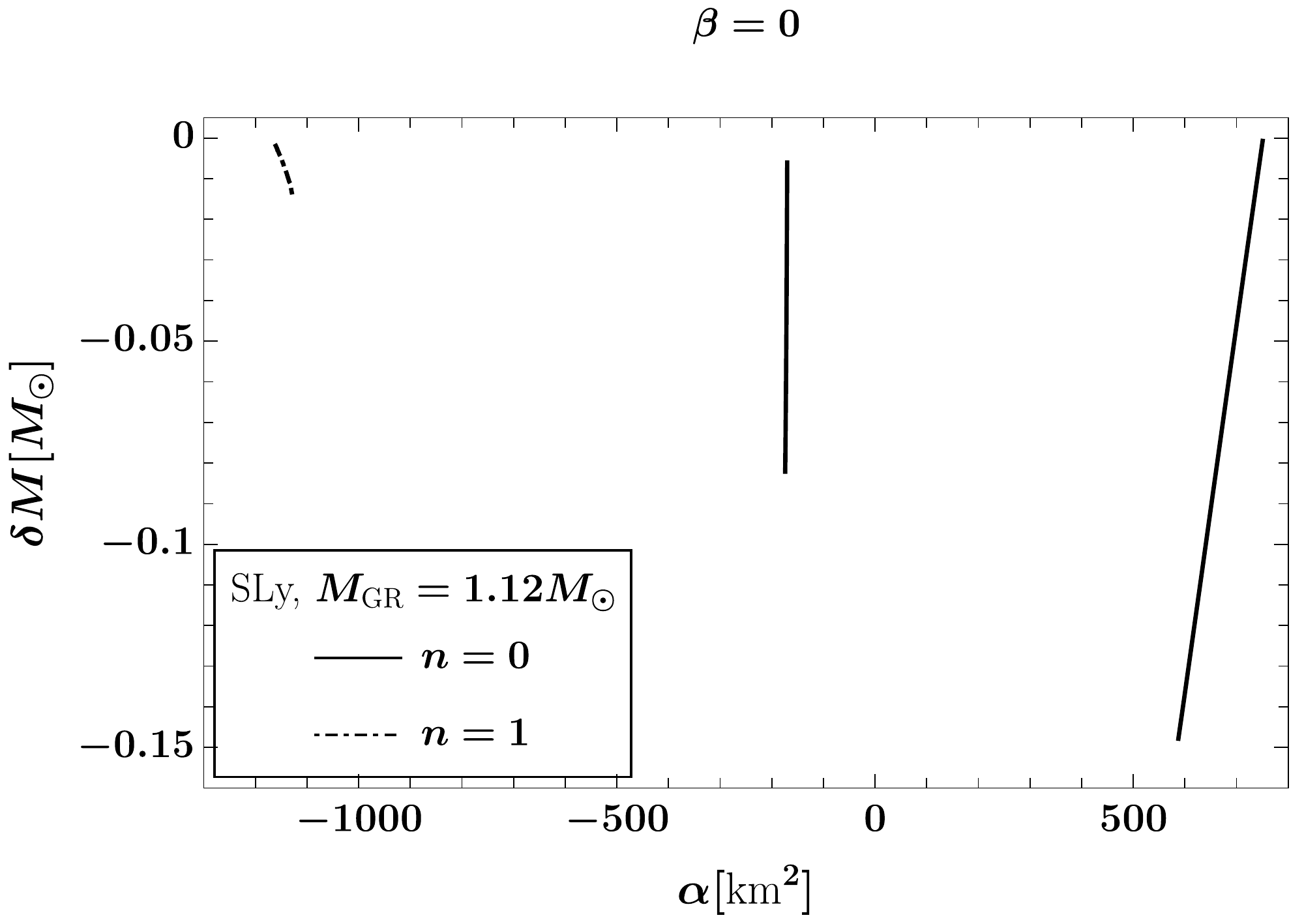}}
	\\
	\subfloat{\includegraphics[width=0.75\linewidth]{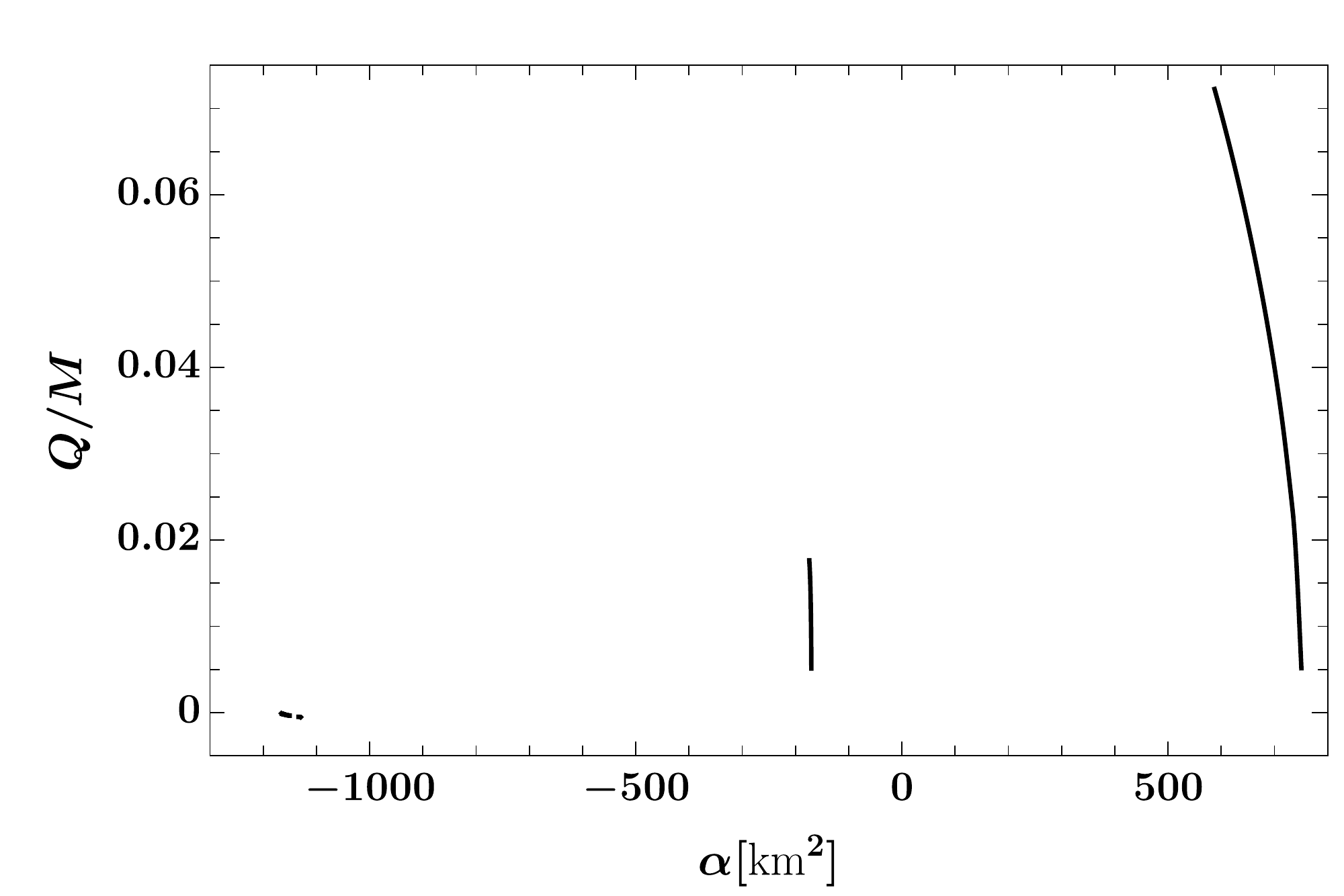}}      
	\caption{Mass difference and scalar charge of the scalarized solutions along the instability lines, for $\beta=0$. The scenario considered here corresponds to $\epsilon_0=8.1\times 10^{17}~\text{kg}/\text{m}^3$ ($M_\text{GR}=1.12M_\odot$) together with the SLy EOS. Solutions with zero node acquire a significant charge and mass difference, and are apparently disconnected from GR when they appear while increasing $\alpha$ towards positive values. Solutions with $n=1$ nodes are very close to GR, with a small charge and mass difference. Since they extend only over a small range of $Q$ and $\delta M$, they are difficult to spot. They lie at the upper left (respectively lower left) of the top (respectively bottom) panel.
	}
	\label{fig:beta0}
\end{figure}

As we mentioned at the end of Sec.~\ref{sec:lightSLy}, a generic feature that is not observable in Figs.~\ref{fig:Sly112}, \ref{fig:MPA1} and \ref{fig:SLy204}, is that scalarized solutions are present in a tiny band close to each instability line.
Let us illustrate this with the light star model (with SLy EOS), that is the one which corresponds to Fig.~\ref{fig:Sly112}. For simplicity, we also restrict our study to solutions with $\beta=0$ (\textit{i.e.}, we take a cut along the vertical axis in Fig.~\ref{fig:Sly112}). The characteristics of the solutions are shown in Fig.~\ref{fig:beta0}.
Scalarized solutions with zero nodes (the ones lying close to the $n=0$ instability line of the GR solution) have a characteristic mass difference and scalar charge which is not particularly small. It is of the same order as for the solutions we previously examined (Figs.~\ref{fig:smallCoup}--\ref{fig:50M204}). They also exhibit a surprising behaviour: when increasing $\alpha$ progressively from 0 towards positive values, the mass and scalar charge suddenly deviate from GR, instead of being smoothly connected; further increasing $\alpha$, $\delta M$ and $Q$ then tend to decrease. This behaviour is significantly different from what we could observe in Figs.~\ref{fig:smallCoup}--\ref{fig:50M204}.
 
 Solutions with more nodes ($n=1$, 2, 3...) exhibit a clear feature: they deviate very slightly from GR in terms of mass, and acquire only a small scalar charge (typically $\delta M < 10^{-2}$ and $Q/M < 10^{-4}$). We verified this behaviour for all higher nodes admitted; however, for simplicity, in Fig.~\ref{fig:beta0} we show only the case $n=1$. This feature can be understood as follows; close to some instability line (on the unstable side), an unstable mode of the effective potential associated with the GR solution has just appeared. A very small deformation of the potential can therefore easily restore the equilibrium. This deformation can be caused by the back-reaction of the scalar onto the metric: the instability is triggered, the scalar field starts growing, but it immediately back-reacts on the potential, making it shallower and suppressing the instability. Clearly, such a behaviour can only happen close to instability lines, where a specific mode is on the edge of stability.

\subsection{Predicting the scalar profile of scalarized stars from GR solutions}\label{Sec:EffMass}

We will conclude this study by arguing that, already at the perturbative level of the GR solution, we can identify an influence on the profile of the scalar field in the fully scalarized solution. To this end, let us focus on the effective mass given in Eq.~\eqref{eq:eff_masss}, $ m_\text{eff}^2=\beta R/2-\alpha \mathscr{G}$. This is a radially dependent quantity, and the scalar field is most likely to grow at radii where $m_\text{eff}^2$ is most negative. In particular, it is natural to expect that, if $m_\text{eff}^2$ has a minimum at $r=0$, this will favor a monotonic profile for the scalar field, and hence an $n=0$ type of solution. On the contrary, if $m_\text{eff}^2$ has a minimum at $r>0$, this favors a peaked profile for the scalar field, which is more common in $n\geq1$ solutions. Let us illustrate this with a concrete example. We will
consider the scenario that corresponds to $M_\text{GR}=1.12 M_\odot$, together with the SLy EOS, and two choices of $\beta$: $\beta=-10$ and $\beta=-100$. In the first case, only solutions with 0 node exist; in the second case, we can construct solutions with 0 or 1 node.

\begin{figure}[ht]
	\subfloat{\includegraphics[width=0.7\linewidth]{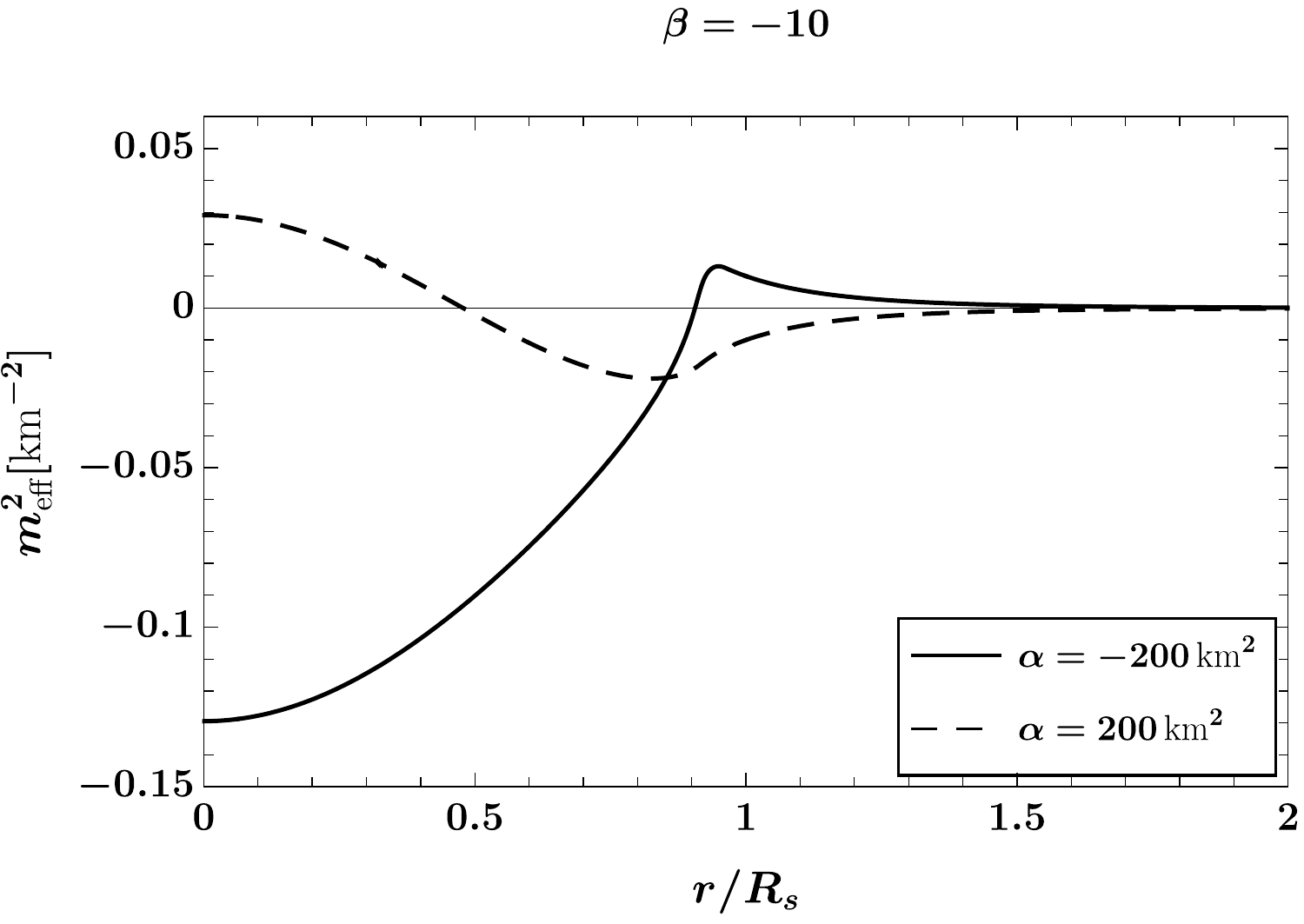}}
	\\%
	\subfloat{\includegraphics[width=0.7\linewidth]{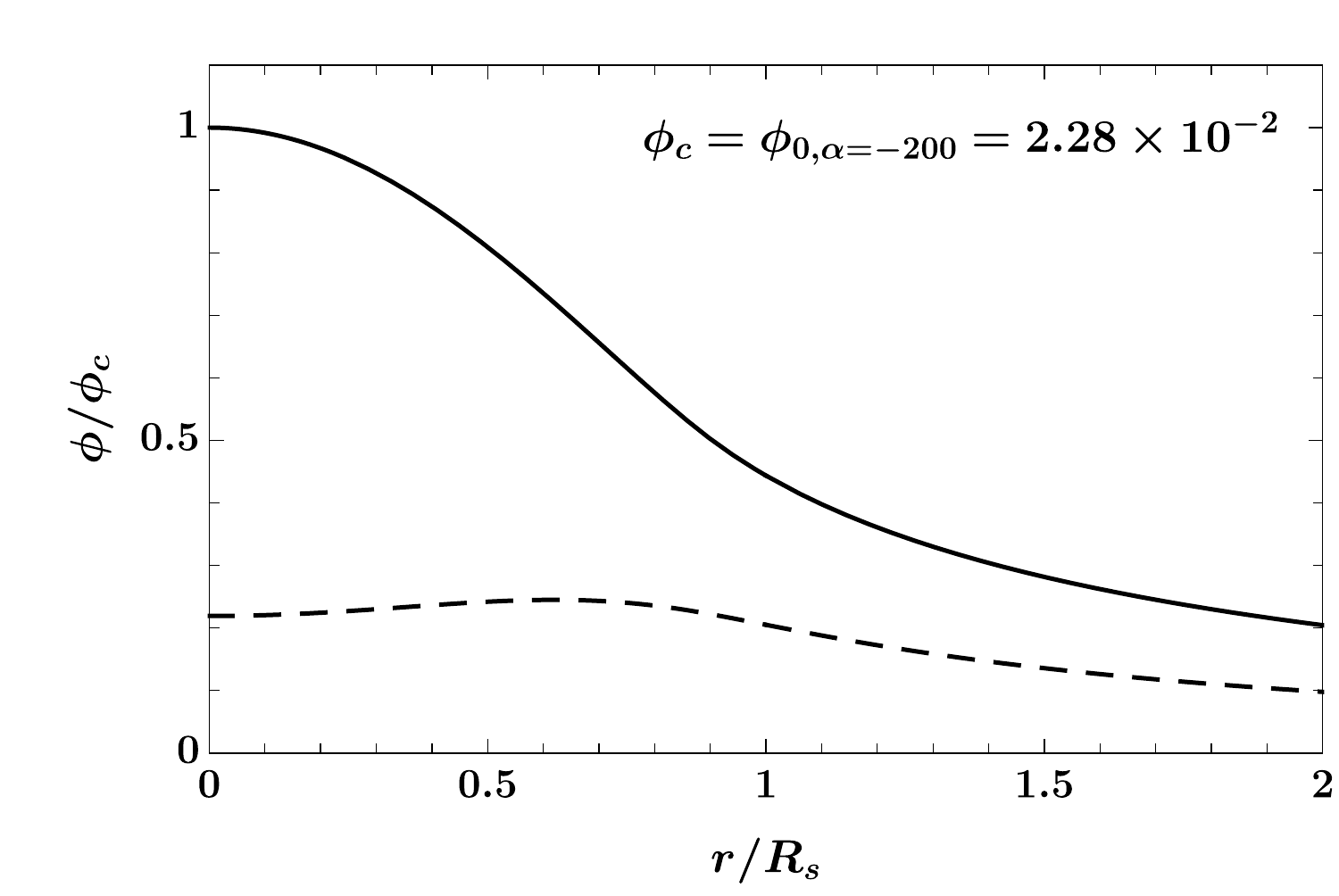}}
	\caption{Upper panel: radial profile of the effective mass squared over the GR background, using the SLy EOS and a central density $\epsilon_0=8.1\times 10^{17}\,\text{kg}/\text{m}^3$ (yielding $M_{\text{GR}}=1.12 M_{\odot}$), for $\beta=-10$ and $\alpha=\pm200$~km$^2$; Lower panel: radial profile of the scalar field, this time in the fully scalarized solution with the same EOS, central density, and Lagrangian parameters. The radial coordinate is normalized by $R_\text{s}$, the radius of the star surface. In the lower panel, the scalar field is normalized to its central value for $\alpha=-200\, \text{km}^2$. When the minimum of $m_\text{eff}^2$ is shifted to $r>0$, so is the peak of $\phi$.}
	\label{fig:EffMass2}
\end{figure}

We first focus on the case $\beta=-10$. The Ricci scalar is everywhere positive over the background we consider, with a maximum at $r=0$; hence, $\beta R$ contributes negatively to the squared mass, favouring the growth of the scalar field close to the center. The Gauss-Bonnet scalar, on the other hand, is negative in the central region of the star, and becomes positive towards the surface. Therefore, $-\alpha\mathscr{G}$ reinforces the effect of $\beta R$ if $\alpha<0$, while couterbalancing it if $\alpha>0$. This is illustrated in the top panel of Fig.~\ref{fig:EffMass2}.
The bottom panel shows the scalar profile of the fully scalarized solutions associated with the same parameters. In this range of parameters, only solutions with 0 node are allowed (as one can check in Fig.~\ref{fig:Sly112}); hence, pushing the minimum of $m_\text{eff}^2$ away from the center cannot favour $n=1$ solutions, which do not exist. Still, we notice that positive $\alpha$ values, which have the effect of displacing the minimum of $m_\text{eff}^2$ to $r>0$, also displace the peak of the scalar field to $r>0$. The peak of the scalar field is located approximately at the minimum of $m_\text{eff}^2$. Again, one must be careful in the comparison of the two panels, as one of them corresponds to a GR star while the other one corresponds to a scalarized star. However, our analysis seems to capture what happens during the transition from the GR to the scalarized branch.

\begin{figure}[ht]
	\subfloat{\includegraphics[width=0.735\linewidth]{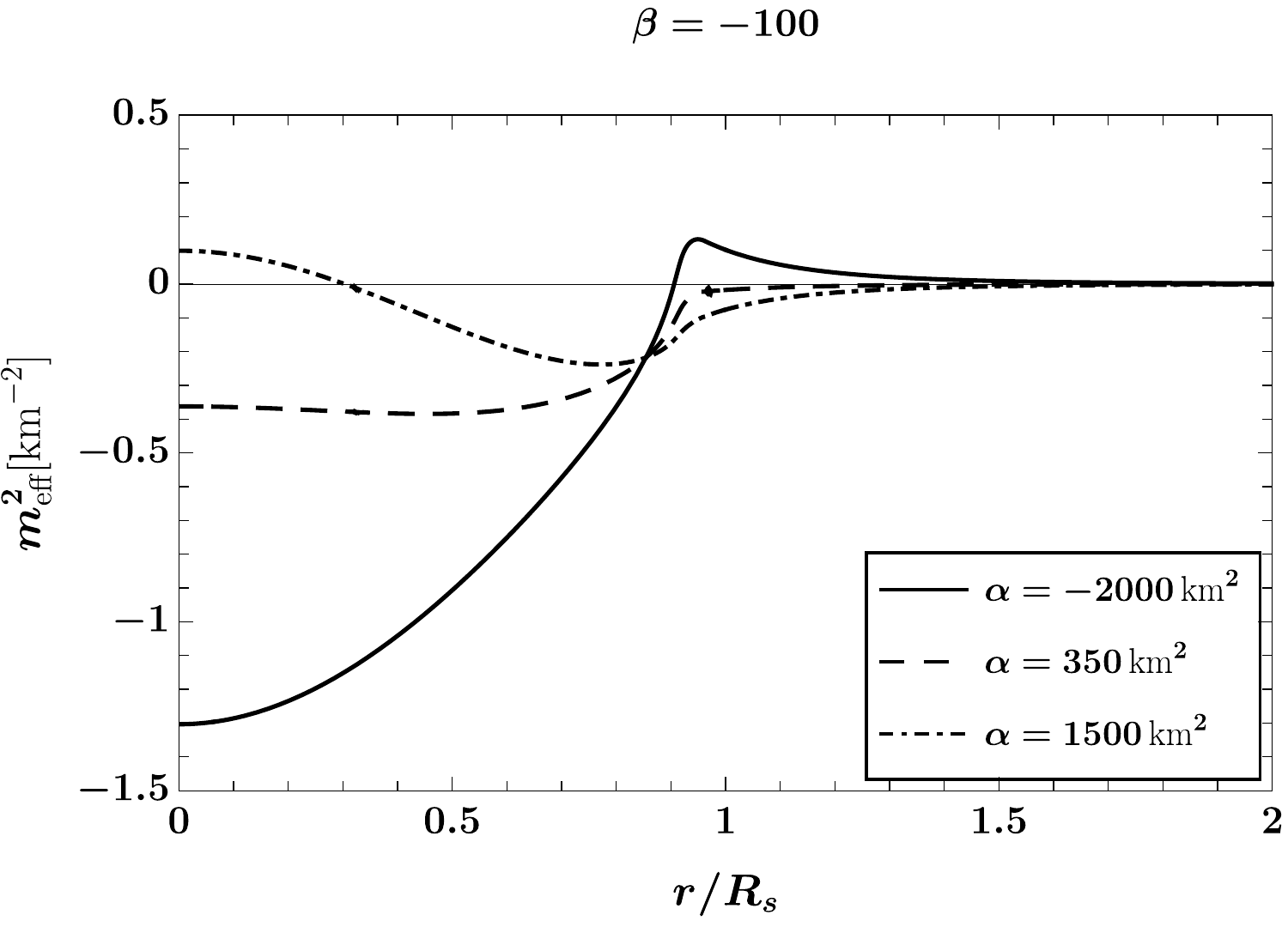}}
		\\%
	\subfloat{\includegraphics[width=0.735\linewidth]{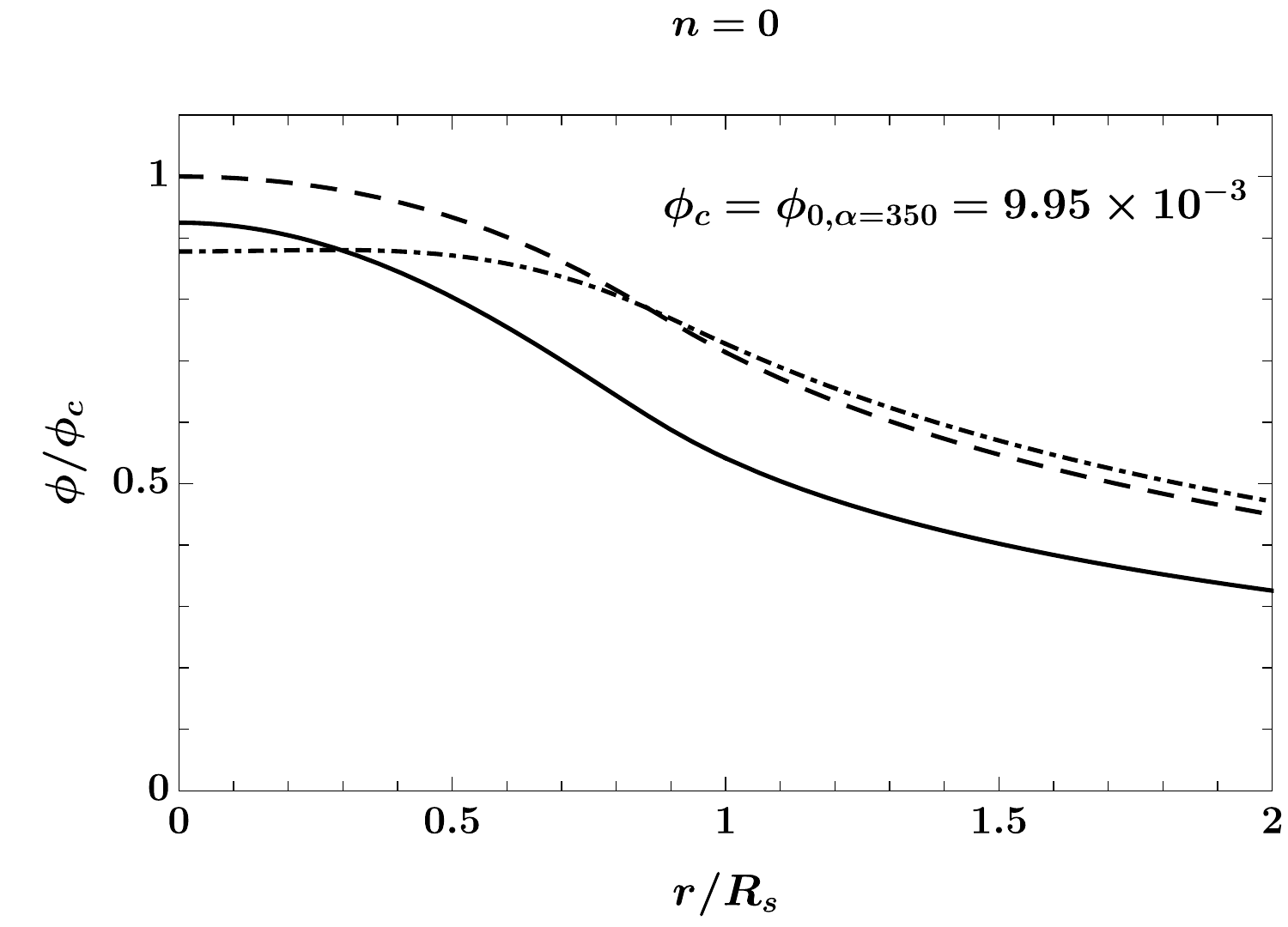}}
	    \\%
	\subfloat{\includegraphics[width=0.735\linewidth]{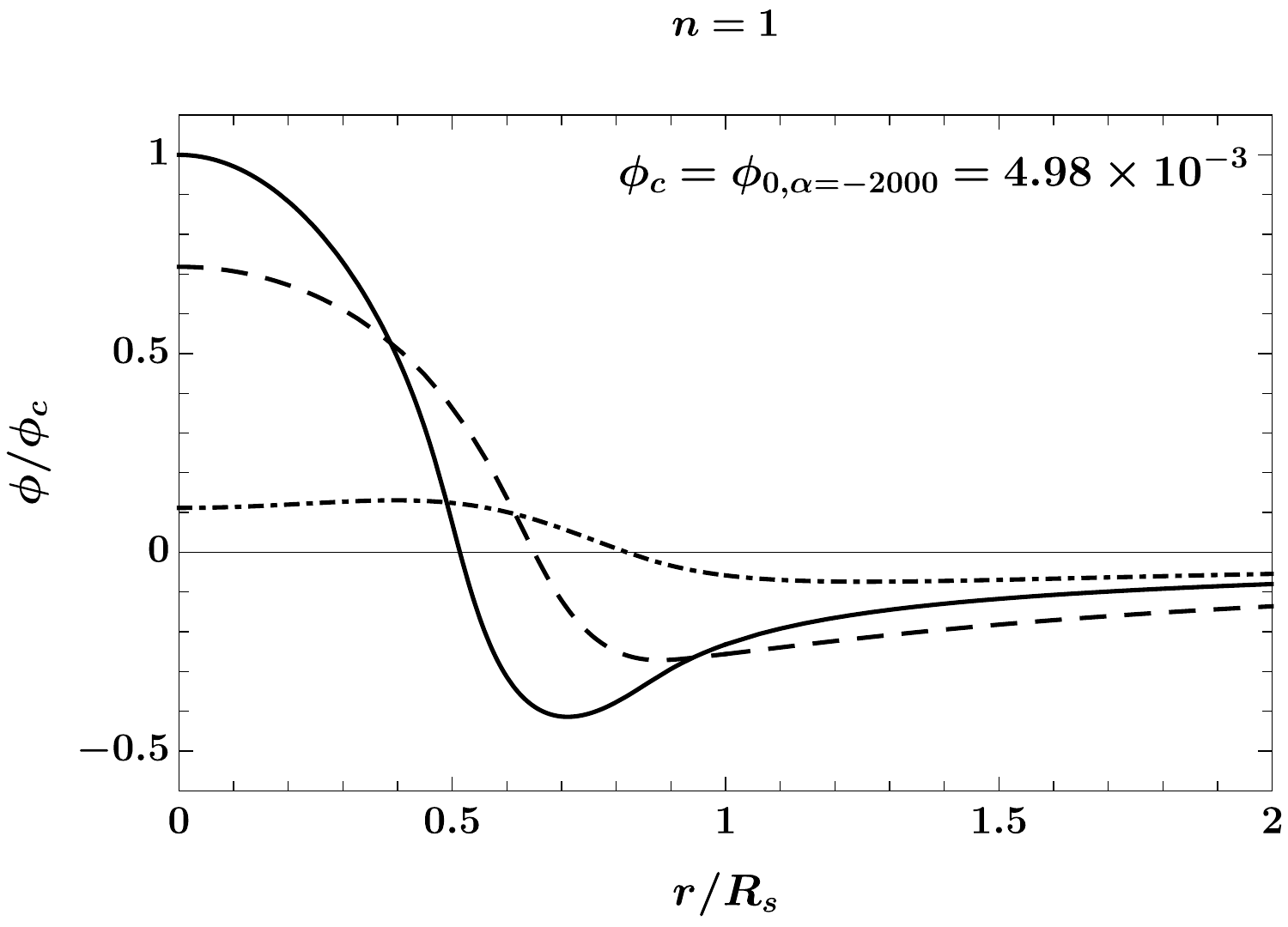}}
	\caption{Upper panel: radial profile of the effective mass squared over the GR background, using the SLy EOS and a central density $\epsilon_0=8.1\times 10^{17}\,\text{kg}/\text{m}^3$ (yielding $M_{\text{GR}}=1.12 M_{\odot}$), for $\beta=-100$ and $\alpha=-200$, 350 or 1500~km$^2$; Center (respectively lower) panel: radial profile of the scalar field solution with 0 (respectively 1) node in the fully scalarized solution with the same EOS, central density, and Lagrangian parameters. The normalization is similar to the one of Fig.~\ref{fig:EffMass2}. When increasing $\alpha$, the minimum of $m_\text{eff}^2$ is progressively shifted from $r=0$ to a finite radius, alternatively favoring the growth of $n=0$ and $n=1$ solutions.}
	\label{fig:EffMass3}
\end{figure}

To illustrate better the transition between $n=0$ and $n=1$ solutions, let us now consider the case $\beta=-100$. 
The qualitative discussion about the effect of $\beta R$ and $-\alpha\mathscr{G}$ over the effective mass is exactly the same as in the previous case. We will therefore consider again a large negative and a large positive value of $\alpha$, as well as an intermediate one: $\alpha=-2000, \,350$ and 1500~km$^2$. Note that the intermediate value corresponds to $\alpha_\text{c}$ in Sec.~\ref{Sec:betaNeg}, the critical value at which scalarized stars with $n=0$ node become more massive (and hence probably less stable) than those with $n=1$ node. We show the results in Fig.~\ref{fig:EffMass3}.
The top panel shows the profile of the effective scalar mass. It behaves exactly as in the case $\beta=-10$, with a minimum at $r=0$ for negative values of $\alpha$, which is progressively shifted to larger radii when we increase $\alpha$. For the parameters we chose, this time, both solutions with zero and one node exist. In the center (respectively bottom)  panel of Fig.~\ref{fig:EffMass3}, we show the $n=0$ (respectively $n=1$) solutions. In Sec.~\ref{Sec:betaNeg}, we stated that for $\alpha<\alpha_c$ we expected that the zero node solution will be energetically preferred over the one node solution, and vice-versa for $\alpha>\alpha_c$. The profiles of the effective mass squared give a complementary argument that strengthens this expectation. Indeed, for $\alpha=-2000\,\text{km}^2\ll\alpha_c$ the shape of $m_\text{eff}^2$ favours a scalar solution with a maximum at the center of the star, which decays monotonically with $r$, \textit{i.e.} a $n=0$ solution. For $\alpha=1500\,\text{km}^2\gg\alpha_c$, the tachyonic instability is still triggered inside the star, but away from the center. Thus, we expect that a solution with one node will be favoured. The transition between a minimum at $r=0$ and $r>0$ indeed seems to occur around $\alpha_\text{c}$.

\section{Conclusions}
\label{sec:discussion}

We have explored scalarized neutron stars when couplings between the scalar field and both the Ricci and the Gauss-Bonnet invariants are present. This completes the analysis initiated in~\cite{Andreou:2019ikc,Ventagli:2020rnx}, where all the terms contributing to the onset of scalarization were identified, and continued in~\cite{Antoniou:2021zoy} with the study of scalarized black holes in this minimal setup.

We have identified the regions of parameter space where solutions exist, considering three different stellar scenarios which correspond to different central densities and EOS. Although we have considered only a limited number of different central densities, we have selected the ones that correspond to the lowest/largest neutron star mass in GR, in order to cover very different setups. The regions where scalarized solutions exist are systematically smaller than the ones where the GR branch is tachyonically unstable. The complementary regions, where the GR solution is unstable while no scalarized solution exists, should be excluded.

We then investigated in detail the physical characteristics of the scalarized solutions. In general, large parameters ($|\beta|\gg1$ or $|\alpha|\gg L^2$, where $L\simeq10$~km is the typical curvature scale) lead to scalar charges that would be in conflict with binary pulsar constraints. However, it is interesting to notice that solutions with $\beta>0$ and reasonably small $\alpha$ (typically $|\alpha|\lesssim 50$~km$^2$) lead either to stable GR configurations, or to scalarized stars with small charges. Remarkably, this is the region of the $(\alpha,\beta)$ parameter space for which GR is a cosmological attractor \cite{Antoniou:2020nax} and black holes scalarization can take place \cite{Antoniou:2021zoy}. Therefore, it is possible to construct scalarization models that are consistent with current observations, while still having interesting strong field phenomenology. It's worth noting that future gravitational-wave observations, such as for instance the observations of extreme mass ratio inspirals by LISA~\cite{Maselli:2020zgv, Maselli:2021men}, will reach the precision to measure small scalar charges for neutron stars and black holes.

We have also discovered that scalarized solutions systematically exist near the thresholds that delimit the stability of the GR solutions, and provided a putative explanation for this. Finally, we have shown that the profile of the effective mass at the GR level can foster the growth of certain modes with respect to others.

An obvious continuation of the present work is the stability analysis of the scalarized solutions, both the neutron stars presented here and the black holes investigated in~\cite{Antoniou:2021zoy}.
It will also be interesting to combine the bounds coming from neutron star and black hole observations with the theoretical constraints that relate to the requirement that scalarization models have a well-posed initial value problem \cite{Ripley:2020vpk}. So far, the combined theory with both Ricci and Gauss-Bonnet couplings has not been studied in detail from the initial value problem perspective. Finally, rotation is known to have important effects on black hole scalarization with a Gauss-Bonnet coupling, either quenching it (for $\alpha>0$ \cite{Cunha:2019dwb,Collodel:2019kkx}) or triggering it (for $\alpha<0$ \cite{Dima:2020yac}). The effect of rotation on neutron star scalarization was investigated in the framework of the DEF model \cite{Doneva:2013qva}. It would be interesting to extend this analysis to coupled Ricci/Gauss-Bonnet couplings, or pure Gauss-Bonnet ones.

\begin{acknowledgments}
 G.A. acknowledges partial support from
the Onassis Foundation.
This project has received funding from the European Union's Horizon 2020 research and innovation programme under the Marie Sklodowska-Curie grant agreement No 101007855.
A.L. thanks FCT for financial support through Project~No.~UIDB/00099/2020.
A.L. acknowledges financial support provided by FCT/Portugal through grants PTDC/MAT-APL/30043/2017 and PTDC/FIS-AST/7002/2020.
T.P.S. acknowledges partial support from the STFC Consolidated Grants No. ST/T000732/1 and No. ST/V005596/1. 
We also acknowledge  networking support by the GWverse COST Action
CA16104, ``Black holes, gravitational waves and fundamental physics.''
\end{acknowledgments}

\appendix
\section{Equations of motion}
We report here the field equations for action~\eqref{eq:ActionCaseI}, where we set $\gamma=0$ and $m_\phi=0$, for a static and spherically symmetric spacetime and with matter described as a perfect fluid:
\begin{widetext}
\begin{align}
\begin{split}
    \underline{\boldsymbol{tt}}:\quad &e^{2 \Lambda}(\beta  \kappa  \phi ^2+2 \kappa  r^2 \epsilon -2)
    +e^{\Lambda}(-8 \alpha  \kappa  \phi  \Lambda ' \phi '+16 \alpha  \kappa  \phi '^2+16 \alpha  \kappa  \phi  \phi ''-\beta  \kappa  \phi ^2+\beta  \kappa  r^2 \phi  \Lambda ' \phi '-2 \beta  \kappa  r^2 \phi '^2\\
    &-2 \beta  \kappa  r^2 \phi  \phi ''+\kappa  r^2 \phi '^2+\beta  \kappa  r \phi ^2 \Lambda '-4 \beta 
   \kappa  r \phi  \phi '-2 r \Lambda '+2)+
   24 \alpha  \kappa  \phi  \Lambda ' \phi '-16 \alpha  \kappa  \phi '^2-16 \alpha  \kappa  \phi  \phi ''=0,
\end{split}\\ \nonumber\\
\begin{split}
    \underline{\boldsymbol{rr}}:\quad &e^{2 \Lambda} (\beta  \kappa  \phi ^2-2 \kappa  p r^2-2)
    +e^\Lambda (8 \alpha  \kappa  \phi  \Gamma' \phi '-\beta  \kappa  r^2 \phi  \Gamma' \phi '-\beta  \kappa  r \phi ^2 \Gamma'\\&+2 r \Gamma'-\beta  \kappa  \phi ^2-\kappa  r^2 \phi '^2-4 \beta  \kappa  r \phi  \phi '+2)-24 \alpha  \kappa  \phi  \Gamma' \phi '=0,
\end{split}\\ \nonumber\\
\begin{split}
    \underline{\textbf{Scalar}}:\quad & 4 \beta   \phi\,e^{2 \Lambda } +e^{\Lambda }(-8 \alpha   \phi  \Gamma ' \Lambda '+8 \alpha   \phi  \Gamma '^2+16 \alpha   \phi  \Gamma ''-4 \beta  \phi
   +\beta   r^2 \phi  \Gamma ' \Lambda '-\beta   r^2 \phi  \Gamma '^2\\
   &-2 \beta   r^2 \phi  \Gamma ''-2   r^2 \Gamma ' \phi '+2  r^2 \Lambda ' \phi '-4   r^2 \phi ''-4 \beta   r \phi  \Gamma '+4 \beta   r \phi  \Lambda '-8  r \phi ')\\
   &+24 \alpha   \phi  \Gamma ' \Lambda '-8 \alpha   \phi (\Gamma '^2+16 \alpha  \phi  \Gamma '')=0,
\end{split}\\ \nonumber\\
\begin{split}
    \underline{\boldsymbol{T_{(m),\mu}^{\mu\nu}}}:\quad &2p'+(\epsilon+p)\Gamma'=0.
\end{split}
\end{align}
\end{widetext}

\bibliography{bibnote}

\begin{thebibliography}{48}%
\makeatletter
\providecommand \@ifxundefined [1]{%
 \@ifx{#1\undefined}
}%
\providecommand \@ifnum [1]{%
 \ifnum #1\expandafter \@firstoftwo
 \else \expandafter \@secondoftwo
 \fi
}%
\providecommand \@ifx [1]{%
 \ifx #1\expandafter \@firstoftwo
 \else \expandafter \@secondoftwo
 \fi
}%
\providecommand \natexlab [1]{#1}%
\providecommand \enquote  [1]{``#1''}%
\providecommand \bibnamefont  [1]{#1}%
\providecommand \bibfnamefont [1]{#1}%
\providecommand \citenamefont [1]{#1}%
\providecommand \href@noop [0]{\@secondoftwo}%
\providecommand \href [0]{\begingroup \@sanitize@url \@href}%
\providecommand \@href[1]{\@@startlink{#1}\@@href}%
\providecommand \@@href[1]{\endgroup#1\@@endlink}%
\providecommand \@sanitize@url [0]{\catcode `\\12\catcode `\$12\catcode
  `\&12\catcode `\#12\catcode `\^12\catcode `\_12\catcode `\%12\relax}%
\providecommand \@@startlink[1]{}%
\providecommand \@@endlink[0]{}%
\providecommand \url  [0]{\begingroup\@sanitize@url \@url }%
\providecommand \@url [1]{\endgroup\@href {#1}{\urlprefix }}%
\providecommand \urlprefix  [0]{URL }%
\providecommand \Eprint [0]{\href }%
\providecommand \doibase [0]{https://doi.org/}%
\providecommand \selectlanguage [0]{\@gobble}%
\providecommand \bibinfo  [0]{\@secondoftwo}%
\providecommand \bibfield  [0]{\@secondoftwo}%
\providecommand \translation [1]{[#1]}%
\providecommand \BibitemOpen [0]{}%
\providecommand \bibitemStop [0]{}%
\providecommand \bibitemNoStop [0]{.\EOS\space}%
\providecommand \EOS [0]{\spacefactor3000\relax}%
\providecommand \BibitemShut  [1]{\csname bibitem#1\endcsname}%
\let\auto@bib@innerbib\@empty
\bibitem [{\citenamefont {Abbott}\ \emph {et~al.}(2016)\citenamefont {Abbott}
  \emph {et~al.}}]{Abbott:2016blz}%
  \BibitemOpen
  \bibfield  {author} {\bibinfo {author} {\bibfnamefont {B.~P.}\ \bibnamefont
  {Abbott}} \emph {et~al.} (\bibinfo {collaboration} {LIGO Scientific,
  Virgo}),\ }\bibfield  {title} {\bibinfo {title} {{Observation of
  Gravitational Waves from a Binary Black Hole Merger}},\ }\href
  {https://doi.org/10.1103/PhysRevLett.116.061102} {\bibfield  {journal}
  {\bibinfo  {journal} {Phys. Rev. Lett.}\ }\textbf {\bibinfo {volume} {116}},\
  \bibinfo {pages} {061102} (\bibinfo {year} {2016})},\ \Eprint
  {https://arxiv.org/abs/1602.03837} {arXiv:1602.03837 [gr-qc]} \BibitemShut
  {NoStop}%
\bibitem [{\citenamefont {Abbott}\ \emph {et~al.}(2017)\citenamefont {Abbott}
  \emph {et~al.}}]{TheLIGOScientific:2017qsa}%
  \BibitemOpen
  \bibfield  {author} {\bibinfo {author} {\bibfnamefont {B.~P.}\ \bibnamefont
  {Abbott}} \emph {et~al.} (\bibinfo {collaboration} {LIGO Scientific,
  Virgo}),\ }\bibfield  {title} {\bibinfo {title} {{GW170817: Observation of
  Gravitational Waves from a Binary Neutron Star Inspiral}},\ }\href
  {https://doi.org/10.1103/PhysRevLett.119.161101} {\bibfield  {journal}
  {\bibinfo  {journal} {Phys. Rev. Lett.}\ }\textbf {\bibinfo {volume} {119}},\
  \bibinfo {pages} {161101} (\bibinfo {year} {2017})},\ \Eprint
  {https://arxiv.org/abs/1710.05832} {arXiv:1710.05832 [gr-qc]} \BibitemShut
  {NoStop}%
\bibitem [{\citenamefont {Abbott}\ \emph {et~al.}(2020)\citenamefont {Abbott}
  \emph {et~al.}}]{Abbott:2020niy}%
  \BibitemOpen
  \bibfield  {author} {\bibinfo {author} {\bibfnamefont {R.}~\bibnamefont
  {Abbott}} \emph {et~al.} (\bibinfo {collaboration} {LIGO Scientific,
  Virgo}),\ }\bibfield  {title} {\bibinfo {title} {{GWTC-2: Compact Binary
  Coalescences Observed by LIGO and Virgo During the First Half of the Third
  Observing Run}},\ }\href@noop {} {\  (\bibinfo {year} {2020})},\ \Eprint
  {https://arxiv.org/abs/2010.14527} {arXiv:2010.14527 [gr-qc]} \BibitemShut
  {NoStop}%
\bibitem [{\citenamefont {Damour}\ and\ \citenamefont
  {Nordtvedt}(1993)}]{Damour:1992kf}%
  \BibitemOpen
  \bibfield  {author} {\bibinfo {author} {\bibfnamefont {T.}~\bibnamefont
  {Damour}}\ and\ \bibinfo {author} {\bibfnamefont {K.}~\bibnamefont
  {Nordtvedt}},\ }\bibfield  {title} {\bibinfo {title} {{General relativity as
  a cosmological attractor of tensor scalar theories}},\ }\href
  {https://doi.org/10.1103/PhysRevLett.70.2217} {\bibfield  {journal} {\bibinfo
   {journal} {Phys. Rev. Lett.}\ }\textbf {\bibinfo {volume} {70}},\ \bibinfo
  {pages} {2217} (\bibinfo {year} {1993})}\BibitemShut {NoStop}%
\bibitem [{\citenamefont {Damour}\ and\ \citenamefont
  {Esposito-Far\`ese}(1993)}]{Damour:1993hw}%
  \BibitemOpen
  \bibfield  {author} {\bibinfo {author} {\bibfnamefont {T.}~\bibnamefont
  {Damour}}\ and\ \bibinfo {author} {\bibfnamefont {G.}~\bibnamefont
  {Esposito-Far\`ese}},\ }\bibfield  {title} {\bibinfo {title}
  {{Nonperturbative strong field effects in tensor - scalar theories of
  gravitation}},\ }\href {https://doi.org/10.1103/PhysRevLett.70.2220}
  {\bibfield  {journal} {\bibinfo  {journal} {Phys. Rev. Lett.}\ }\textbf
  {\bibinfo {volume} {70}},\ \bibinfo {pages} {2220} (\bibinfo {year}
  {1993})}\BibitemShut {NoStop}%
\bibitem [{\citenamefont {Hawking}(1972)}]{Hawking:1972qk}%
  \BibitemOpen
  \bibfield  {author} {\bibinfo {author} {\bibfnamefont {S.~W.}\ \bibnamefont
  {Hawking}},\ }\bibfield  {title} {\bibinfo {title} {{Black holes in the
  Brans-Dicke theory of gravitation}},\ }\href
  {https://doi.org/10.1007/BF01877518} {\bibfield  {journal} {\bibinfo
  {journal} {Commun. Math. Phys.}\ }\textbf {\bibinfo {volume} {25}},\ \bibinfo
  {pages} {167} (\bibinfo {year} {1972})}\BibitemShut {NoStop}%
\bibitem [{\citenamefont {Sotiriou}\ and\ \citenamefont
  {Faraoni}(2012)}]{Sotiriou:2011dz}%
  \BibitemOpen
  \bibfield  {author} {\bibinfo {author} {\bibfnamefont {T.~P.}\ \bibnamefont
  {Sotiriou}}\ and\ \bibinfo {author} {\bibfnamefont {V.}~\bibnamefont
  {Faraoni}},\ }\bibfield  {title} {\bibinfo {title} {{Black holes in
  scalar-tensor gravity}},\ }\href
  {https://doi.org/10.1103/PhysRevLett.108.081103} {\bibfield  {journal}
  {\bibinfo  {journal} {Phys. Rev. Lett.}\ }\textbf {\bibinfo {volume} {108}},\
  \bibinfo {pages} {081103} (\bibinfo {year} {2012})},\ \Eprint
  {https://arxiv.org/abs/1109.6324} {arXiv:1109.6324 [gr-qc]} \BibitemShut
  {NoStop}%
\bibitem [{\citenamefont {Cardoso}\ \emph
  {et~al.}(2013{\natexlab{a}})\citenamefont {Cardoso}, \citenamefont {Carucci},
  \citenamefont {Pani},\ and\ \citenamefont {Sotiriou}}]{Cardoso:2013opa}%
  \BibitemOpen
  \bibfield  {author} {\bibinfo {author} {\bibfnamefont {V.}~\bibnamefont
  {Cardoso}}, \bibinfo {author} {\bibfnamefont {I.~P.}\ \bibnamefont
  {Carucci}}, \bibinfo {author} {\bibfnamefont {P.}~\bibnamefont {Pani}},\ and\
  \bibinfo {author} {\bibfnamefont {T.~P.}\ \bibnamefont {Sotiriou}},\
  }\bibfield  {title} {\bibinfo {title} {{Matter around Kerr black holes in
  scalar-tensor theories: scalarization and superradiant instability}},\ }\href
  {https://doi.org/10.1103/PhysRevD.88.044056} {\bibfield  {journal} {\bibinfo
  {journal} {Phys. Rev.}\ }\textbf {\bibinfo {volume} {D88}},\ \bibinfo {pages}
  {044056} (\bibinfo {year} {2013}{\natexlab{a}})},\ \Eprint
  {https://arxiv.org/abs/1305.6936} {arXiv:1305.6936 [gr-qc]} \BibitemShut
  {NoStop}%
\bibitem [{\citenamefont {Cardoso}\ \emph
  {et~al.}(2013{\natexlab{b}})\citenamefont {Cardoso}, \citenamefont {Carucci},
  \citenamefont {Pani},\ and\ \citenamefont {Sotiriou}}]{Cardoso:2013fwa}%
  \BibitemOpen
  \bibfield  {author} {\bibinfo {author} {\bibfnamefont {V.}~\bibnamefont
  {Cardoso}}, \bibinfo {author} {\bibfnamefont {I.~P.}\ \bibnamefont
  {Carucci}}, \bibinfo {author} {\bibfnamefont {P.}~\bibnamefont {Pani}},\ and\
  \bibinfo {author} {\bibfnamefont {T.~P.}\ \bibnamefont {Sotiriou}},\
  }\bibfield  {title} {\bibinfo {title} {{Black holes with surrounding matter
  in scalar-tensor theories}},\ }\href
  {https://doi.org/10.1103/PhysRevLett.111.111101} {\bibfield  {journal}
  {\bibinfo  {journal} {Phys. Rev. Lett.}\ }\textbf {\bibinfo {volume} {111}},\
  \bibinfo {pages} {111101} (\bibinfo {year} {2013}{\natexlab{b}})},\ \Eprint
  {https://arxiv.org/abs/1308.6587} {arXiv:1308.6587 [gr-qc]} \BibitemShut
  {NoStop}%
\bibitem [{\citenamefont {Palenzuela}\ \emph {et~al.}(2014)\citenamefont
  {Palenzuela}, \citenamefont {Barausse}, \citenamefont {Ponce},\ and\
  \citenamefont {Lehner}}]{Palenzuela:2013hsa}%
  \BibitemOpen
  \bibfield  {author} {\bibinfo {author} {\bibfnamefont {C.}~\bibnamefont
  {Palenzuela}}, \bibinfo {author} {\bibfnamefont {E.}~\bibnamefont
  {Barausse}}, \bibinfo {author} {\bibfnamefont {M.}~\bibnamefont {Ponce}},\
  and\ \bibinfo {author} {\bibfnamefont {L.}~\bibnamefont {Lehner}},\
  }\bibfield  {title} {\bibinfo {title} {{Dynamical scalarization of neutron
  stars in scalar-tensor gravity theories}},\ }\href
  {https://doi.org/10.1103/PhysRevD.89.044024} {\bibfield  {journal} {\bibinfo
  {journal} {Phys. Rev. D}\ }\textbf {\bibinfo {volume} {89}},\ \bibinfo
  {pages} {044024} (\bibinfo {year} {2014})},\ \Eprint
  {https://arxiv.org/abs/1310.4481} {arXiv:1310.4481 [gr-qc]} \BibitemShut
  {NoStop}%
\bibitem [{\citenamefont {Silva}\ \emph {et~al.}(2018)\citenamefont {Silva},
  \citenamefont {Sakstein}, \citenamefont {Gualtieri}, \citenamefont
  {Sotiriou},\ and\ \citenamefont {Berti}}]{Silva:2017uqg}%
  \BibitemOpen
  \bibfield  {author} {\bibinfo {author} {\bibfnamefont {H.~O.}\ \bibnamefont
  {Silva}}, \bibinfo {author} {\bibfnamefont {J.}~\bibnamefont {Sakstein}},
  \bibinfo {author} {\bibfnamefont {L.}~\bibnamefont {Gualtieri}}, \bibinfo
  {author} {\bibfnamefont {T.~P.}\ \bibnamefont {Sotiriou}},\ and\ \bibinfo
  {author} {\bibfnamefont {E.}~\bibnamefont {Berti}},\ }\bibfield  {title}
  {\bibinfo {title} {{Spontaneous scalarization of black holes and compact
  stars from a Gauss-Bonnet coupling}},\ }\href
  {https://doi.org/10.1103/PhysRevLett.120.131104} {\bibfield  {journal}
  {\bibinfo  {journal} {Phys. Rev. Lett.}\ }\textbf {\bibinfo {volume} {120}},\
  \bibinfo {pages} {131104} (\bibinfo {year} {2018})},\ \Eprint
  {https://arxiv.org/abs/1711.02080} {arXiv:1711.02080 [gr-qc]} \BibitemShut
  {NoStop}%
\bibitem [{\citenamefont {Doneva}\ and\ \citenamefont
  {Yazadjiev}(2018{\natexlab{a}})}]{Doneva:2017bvd}%
  \BibitemOpen
  \bibfield  {author} {\bibinfo {author} {\bibfnamefont {D.~D.}\ \bibnamefont
  {Doneva}}\ and\ \bibinfo {author} {\bibfnamefont {S.~S.}\ \bibnamefont
  {Yazadjiev}},\ }\bibfield  {title} {\bibinfo {title} {{New Gauss-Bonnet Black
  Holes with Curvature-Induced Scalarization in Extended Scalar-Tensor
  Theories}},\ }\href {https://doi.org/10.1103/PhysRevLett.120.131103}
  {\bibfield  {journal} {\bibinfo  {journal} {Phys. Rev. Lett.}\ }\textbf
  {\bibinfo {volume} {120}},\ \bibinfo {pages} {131103} (\bibinfo {year}
  {2018}{\natexlab{a}})},\ \Eprint {https://arxiv.org/abs/1711.01187}
  {arXiv:1711.01187 [gr-qc]} \BibitemShut {NoStop}%
\bibitem [{\citenamefont {Doneva}\ and\ \citenamefont
  {Yazadjiev}(2018{\natexlab{b}})}]{Doneva:2017duq}%
  \BibitemOpen
  \bibfield  {author} {\bibinfo {author} {\bibfnamefont {D.~D.}\ \bibnamefont
  {Doneva}}\ and\ \bibinfo {author} {\bibfnamefont {S.~S.}\ \bibnamefont
  {Yazadjiev}},\ }\bibfield  {title} {\bibinfo {title} {{Neutron star solutions
  with curvature induced scalarization in the extended Gauss-Bonnet
  scalar-tensor theories}},\ }\href
  {https://doi.org/10.1088/1475-7516/2018/04/011} {\bibfield  {journal}
  {\bibinfo  {journal} {JCAP}\ }\textbf {\bibinfo {volume} {04}},\ \bibinfo
  {pages} {011}},\ \Eprint {https://arxiv.org/abs/1712.03715} {arXiv:1712.03715
  [gr-qc]} \BibitemShut {NoStop}%
\bibitem [{\citenamefont {Ramazanoğlu}\ and\ \citenamefont
  {Pretorius}(2016)}]{Ramazanoglu:2016kul}%
  \BibitemOpen
  \bibfield  {author} {\bibinfo {author} {\bibfnamefont {F.~M.}\ \bibnamefont
  {Ramazanoğlu}}\ and\ \bibinfo {author} {\bibfnamefont {F.}~\bibnamefont
  {Pretorius}},\ }\bibfield  {title} {\bibinfo {title} {{Spontaneous
  Scalarization with Massive Fields}},\ }\href
  {https://doi.org/10.1103/PhysRevD.93.064005} {\bibfield  {journal} {\bibinfo
  {journal} {Phys. Rev.}\ }\textbf {\bibinfo {volume} {D93}},\ \bibinfo {pages}
  {064005} (\bibinfo {year} {2016})},\ \Eprint
  {https://arxiv.org/abs/1601.07475} {arXiv:1601.07475 [gr-qc]} \BibitemShut
  {NoStop}%
\bibitem [{\citenamefont {Blázquez-Salcedo}\ \emph {et~al.}(2018)\citenamefont
  {Blázquez-Salcedo}, \citenamefont {Doneva}, \citenamefont {Kunz},\ and\
  \citenamefont {Yazadjiev}}]{Blazquez-Salcedo:2018jnn}%
  \BibitemOpen
  \bibfield  {author} {\bibinfo {author} {\bibfnamefont {J.~L.}\ \bibnamefont
  {Blázquez-Salcedo}}, \bibinfo {author} {\bibfnamefont {D.~D.}\ \bibnamefont
  {Doneva}}, \bibinfo {author} {\bibfnamefont {J.}~\bibnamefont {Kunz}},\ and\
  \bibinfo {author} {\bibfnamefont {S.~S.}\ \bibnamefont {Yazadjiev}},\
  }\bibfield  {title} {\bibinfo {title} {{Radial perturbations of the
  scalarized Einstein-Gauss-Bonnet black holes}},\ }\href
  {https://doi.org/10.1103/PhysRevD.98.084011} {\bibfield  {journal} {\bibinfo
  {journal} {Phys. Rev. D}\ }\textbf {\bibinfo {volume} {98}},\ \bibinfo
  {pages} {084011} (\bibinfo {year} {2018})},\ \Eprint
  {https://arxiv.org/abs/1805.05755} {arXiv:1805.05755 [gr-qc]} \BibitemShut
  {NoStop}%
\bibitem [{\citenamefont {Macedo}\ \emph {et~al.}(2019)\citenamefont {Macedo},
  \citenamefont {Sakstein}, \citenamefont {Berti}, \citenamefont {Gualtieri},
  \citenamefont {Silva},\ and\ \citenamefont {Sotiriou}}]{Macedo:2019sem}%
  \BibitemOpen
  \bibfield  {author} {\bibinfo {author} {\bibfnamefont {C.~F.}\ \bibnamefont
  {Macedo}}, \bibinfo {author} {\bibfnamefont {J.}~\bibnamefont {Sakstein}},
  \bibinfo {author} {\bibfnamefont {E.}~\bibnamefont {Berti}}, \bibinfo
  {author} {\bibfnamefont {L.}~\bibnamefont {Gualtieri}}, \bibinfo {author}
  {\bibfnamefont {H.~O.}\ \bibnamefont {Silva}},\ and\ \bibinfo {author}
  {\bibfnamefont {T.~P.}\ \bibnamefont {Sotiriou}},\ }\bibfield  {title}
  {\bibinfo {title} {{Self-interactions and Spontaneous Black Hole
  Scalarization}},\ }\href {https://doi.org/10.1103/PhysRevD.99.104041}
  {\bibfield  {journal} {\bibinfo  {journal} {Phys. Rev. D}\ }\textbf {\bibinfo
  {volume} {99}},\ \bibinfo {pages} {104041} (\bibinfo {year} {2019})},\
  \Eprint {https://arxiv.org/abs/1903.06784} {arXiv:1903.06784 [gr-qc]}
  \BibitemShut {NoStop}%
\bibitem [{\citenamefont {Herdeiro}\ \emph {et~al.}(2018)\citenamefont
  {Herdeiro}, \citenamefont {Radu}, \citenamefont {Sanchis-Gual},\ and\
  \citenamefont {Font}}]{Herdeiro:2018wub}%
  \BibitemOpen
  \bibfield  {author} {\bibinfo {author} {\bibfnamefont {C.~A.}\ \bibnamefont
  {Herdeiro}}, \bibinfo {author} {\bibfnamefont {E.}~\bibnamefont {Radu}},
  \bibinfo {author} {\bibfnamefont {N.}~\bibnamefont {Sanchis-Gual}},\ and\
  \bibinfo {author} {\bibfnamefont {J.~A.}\ \bibnamefont {Font}},\ }\bibfield
  {title} {\bibinfo {title} {{Spontaneous Scalarization of Charged Black
  Holes}},\ }\href {https://doi.org/10.1103/PhysRevLett.121.101102} {\bibfield
  {journal} {\bibinfo  {journal} {Phys. Rev. Lett.}\ }\textbf {\bibinfo
  {volume} {121}},\ \bibinfo {pages} {101102} (\bibinfo {year} {2018})},\
  \Eprint {https://arxiv.org/abs/1806.05190} {arXiv:1806.05190 [gr-qc]}
  \BibitemShut {NoStop}%
\bibitem [{\citenamefont {Ramazanoğlu}(2017)}]{Ramazanoglu:2017xbl}%
  \BibitemOpen
  \bibfield  {author} {\bibinfo {author} {\bibfnamefont {F.~M.}\ \bibnamefont
  {Ramazanoğlu}},\ }\bibfield  {title} {\bibinfo {title} {{Spontaneous growth
  of vector fields in gravity}},\ }\href
  {https://doi.org/10.1103/PhysRevD.96.064009} {\bibfield  {journal} {\bibinfo
  {journal} {Phys. Rev. D}\ }\textbf {\bibinfo {volume} {96}},\ \bibinfo
  {pages} {064009} (\bibinfo {year} {2017})},\ \Eprint
  {https://arxiv.org/abs/1706.01056} {arXiv:1706.01056 [gr-qc]} \BibitemShut
  {NoStop}%
\bibitem [{\citenamefont {Ramazanoğlu}(2018)}]{Ramazanoglu:2018hwk}%
  \BibitemOpen
  \bibfield  {author} {\bibinfo {author} {\bibfnamefont {F.~M.}\ \bibnamefont
  {Ramazanoğlu}},\ }\bibfield  {title} {\bibinfo {title} {{Spontaneous growth
  of spinor fields in gravity}},\ }\href
  {https://doi.org/10.1103/PhysRevD.98.044011} {\bibfield  {journal} {\bibinfo
  {journal} {Phys. Rev. D}\ }\textbf {\bibinfo {volume} {98}},\ \bibinfo
  {pages} {044011} (\bibinfo {year} {2018})},\ \bibinfo {note} {[Erratum:
  Phys.Rev.D 100, 029903 (2019)]},\ \Eprint {https://arxiv.org/abs/1804.00594}
  {arXiv:1804.00594 [gr-qc]} \BibitemShut {NoStop}%
\bibitem [{\citenamefont {Dima}\ \emph {et~al.}(2020)\citenamefont {Dima},
  \citenamefont {Barausse}, \citenamefont {Franchini},\ and\ \citenamefont
  {Sotiriou}}]{Dima:2020yac}%
  \BibitemOpen
  \bibfield  {author} {\bibinfo {author} {\bibfnamefont {A.}~\bibnamefont
  {Dima}}, \bibinfo {author} {\bibfnamefont {E.}~\bibnamefont {Barausse}},
  \bibinfo {author} {\bibfnamefont {N.}~\bibnamefont {Franchini}},\ and\
  \bibinfo {author} {\bibfnamefont {T.~P.}\ \bibnamefont {Sotiriou}},\
  }\bibfield  {title} {\bibinfo {title} {{Spin-induced black hole spontaneous
  scalarization}},\ }\href {https://doi.org/10.1103/PhysRevLett.125.231101}
  {\bibfield  {journal} {\bibinfo  {journal} {Phys. Rev. Lett.}\ }\textbf
  {\bibinfo {volume} {125}},\ \bibinfo {pages} {231101} (\bibinfo {year}
  {2020})},\ \Eprint {https://arxiv.org/abs/2006.03095} {arXiv:2006.03095
  [gr-qc]} \BibitemShut {NoStop}%
\bibitem [{\citenamefont {Herdeiro}\ \emph {et~al.}(2021)\citenamefont
  {Herdeiro}, \citenamefont {Radu}, \citenamefont {Silva}, \citenamefont
  {Sotiriou},\ and\ \citenamefont {Yunes}}]{Herdeiro:2020wei}%
  \BibitemOpen
  \bibfield  {author} {\bibinfo {author} {\bibfnamefont {C.~A.~R.}\
  \bibnamefont {Herdeiro}}, \bibinfo {author} {\bibfnamefont {E.}~\bibnamefont
  {Radu}}, \bibinfo {author} {\bibfnamefont {H.~O.}\ \bibnamefont {Silva}},
  \bibinfo {author} {\bibfnamefont {T.~P.}\ \bibnamefont {Sotiriou}},\ and\
  \bibinfo {author} {\bibfnamefont {N.}~\bibnamefont {Yunes}},\ }\bibfield
  {title} {\bibinfo {title} {{Spin-induced scalarized black holes}},\ }\href
  {https://doi.org/10.1103/PhysRevLett.126.011103} {\bibfield  {journal}
  {\bibinfo  {journal} {Phys. Rev. Lett.}\ }\textbf {\bibinfo {volume} {126}},\
  \bibinfo {pages} {011103} (\bibinfo {year} {2021})},\ \Eprint
  {https://arxiv.org/abs/2009.03904} {arXiv:2009.03904 [gr-qc]} \BibitemShut
  {NoStop}%
\bibitem [{\citenamefont {Berti}\ \emph {et~al.}(2021)\citenamefont {Berti},
  \citenamefont {Collodel}, \citenamefont {Kleihaus},\ and\ \citenamefont
  {Kunz}}]{Berti:2020kgk}%
  \BibitemOpen
  \bibfield  {author} {\bibinfo {author} {\bibfnamefont {E.}~\bibnamefont
  {Berti}}, \bibinfo {author} {\bibfnamefont {L.~G.}\ \bibnamefont {Collodel}},
  \bibinfo {author} {\bibfnamefont {B.}~\bibnamefont {Kleihaus}},\ and\
  \bibinfo {author} {\bibfnamefont {J.}~\bibnamefont {Kunz}},\ }\bibfield
  {title} {\bibinfo {title} {{Spin-induced black-hole scalarization in
  Einstein-scalar-Gauss-Bonnet theory}},\ }\href
  {https://doi.org/10.1103/PhysRevLett.126.011104} {\bibfield  {journal}
  {\bibinfo  {journal} {Phys. Rev. Lett.}\ }\textbf {\bibinfo {volume} {126}},\
  \bibinfo {pages} {011104} (\bibinfo {year} {2021})},\ \Eprint
  {https://arxiv.org/abs/2009.03905} {arXiv:2009.03905 [gr-qc]} \BibitemShut
  {NoStop}%
\bibitem [{\citenamefont {Sotiriou}\ and\ \citenamefont
  {Zhou}(2014{\natexlab{a}})}]{Sotiriou:2013qea}%
  \BibitemOpen
  \bibfield  {author} {\bibinfo {author} {\bibfnamefont {T.~P.}\ \bibnamefont
  {Sotiriou}}\ and\ \bibinfo {author} {\bibfnamefont {S.-Y.}\ \bibnamefont
  {Zhou}},\ }\bibfield  {title} {\bibinfo {title} {{Black hole hair in
  generalized scalar-tensor gravity}},\ }\href
  {https://doi.org/10.1103/PhysRevLett.112.251102} {\bibfield  {journal}
  {\bibinfo  {journal} {Phys. Rev. Lett.}\ }\textbf {\bibinfo {volume} {112}},\
  \bibinfo {pages} {251102} (\bibinfo {year} {2014}{\natexlab{a}})},\ \Eprint
  {https://arxiv.org/abs/1312.3622} {arXiv:1312.3622 [gr-qc]} \BibitemShut
  {NoStop}%
\bibitem [{\citenamefont {Sotiriou}\ and\ \citenamefont
  {Zhou}(2014{\natexlab{b}})}]{Sotiriou:2014pfa}%
  \BibitemOpen
  \bibfield  {author} {\bibinfo {author} {\bibfnamefont {T.~P.}\ \bibnamefont
  {Sotiriou}}\ and\ \bibinfo {author} {\bibfnamefont {S.-Y.}\ \bibnamefont
  {Zhou}},\ }\bibfield  {title} {\bibinfo {title} {{Black hole hair in
  generalized scalar-tensor gravity: An explicit example}},\ }\href
  {https://doi.org/10.1103/PhysRevD.90.124063} {\bibfield  {journal} {\bibinfo
  {journal} {Phys. Rev.}\ }\textbf {\bibinfo {volume} {D90}},\ \bibinfo {pages}
  {124063} (\bibinfo {year} {2014}{\natexlab{b}})},\ \Eprint
  {https://arxiv.org/abs/1408.1698} {arXiv:1408.1698 [gr-qc]} \BibitemShut
  {NoStop}%
\bibitem [{\citenamefont {Antoniou}\ \emph
  {et~al.}(2018{\natexlab{a}})\citenamefont {Antoniou}, \citenamefont
  {Bakopoulos},\ and\ \citenamefont {Kanti}}]{Antoniou:2017acq}%
  \BibitemOpen
  \bibfield  {author} {\bibinfo {author} {\bibfnamefont {G.}~\bibnamefont
  {Antoniou}}, \bibinfo {author} {\bibfnamefont {A.}~\bibnamefont
  {Bakopoulos}},\ and\ \bibinfo {author} {\bibfnamefont {P.}~\bibnamefont
  {Kanti}},\ }\bibfield  {title} {\bibinfo {title} {{Evasion of No-Hair
  Theorems and Novel Black-Hole Solutions in Gauss-Bonnet Theories}},\ }\href
  {https://doi.org/10.1103/PhysRevLett.120.131102} {\bibfield  {journal}
  {\bibinfo  {journal} {Phys. Rev. Lett.}\ }\textbf {\bibinfo {volume} {120}},\
  \bibinfo {pages} {131102} (\bibinfo {year} {2018}{\natexlab{a}})},\ \Eprint
  {https://arxiv.org/abs/1711.03390} {arXiv:1711.03390 [hep-th]} \BibitemShut
  {NoStop}%
\bibitem [{\citenamefont {Antoniou}\ \emph
  {et~al.}(2018{\natexlab{b}})\citenamefont {Antoniou}, \citenamefont
  {Bakopoulos},\ and\ \citenamefont {Kanti}}]{Antoniou:2017hxj}%
  \BibitemOpen
  \bibfield  {author} {\bibinfo {author} {\bibfnamefont {G.}~\bibnamefont
  {Antoniou}}, \bibinfo {author} {\bibfnamefont {A.}~\bibnamefont
  {Bakopoulos}},\ and\ \bibinfo {author} {\bibfnamefont {P.}~\bibnamefont
  {Kanti}},\ }\bibfield  {title} {\bibinfo {title} {{Black-Hole Solutions with
  Scalar Hair in Einstein-Scalar-Gauss-Bonnet Theories}},\ }\href
  {https://doi.org/10.1103/PhysRevD.97.084037} {\bibfield  {journal} {\bibinfo
  {journal} {Phys. Rev.}\ }\textbf {\bibinfo {volume} {D97}},\ \bibinfo {pages}
  {084037} (\bibinfo {year} {2018}{\natexlab{b}})},\ \Eprint
  {https://arxiv.org/abs/1711.07431} {arXiv:1711.07431 [hep-th]} \BibitemShut
  {NoStop}%
\bibitem [{\citenamefont {Doneva}\ and\ \citenamefont
  {Yazadjiev}(2021)}]{Doneva:2021tvn}%
  \BibitemOpen
  \bibfield  {author} {\bibinfo {author} {\bibfnamefont {D.~D.}\ \bibnamefont
  {Doneva}}\ and\ \bibinfo {author} {\bibfnamefont {S.~S.}\ \bibnamefont
  {Yazadjiev}},\ }\bibfield  {title} {\bibinfo {title} {{Beyond the spontaneous
  scalarization: New fully nonlinear dynamical mechanism for formation of
  scalarized black holes}},\ }\href@noop {} {\  (\bibinfo {year} {2021})},\
  \Eprint {https://arxiv.org/abs/2107.01738} {arXiv:2107.01738 [gr-qc]}
  \BibitemShut {NoStop}%
\bibitem [{\citenamefont {Andreou}\ \emph {et~al.}(2019)\citenamefont
  {Andreou}, \citenamefont {Franchini}, \citenamefont {Ventagli},\ and\
  \citenamefont {Sotiriou}}]{Andreou:2019ikc}%
  \BibitemOpen
  \bibfield  {author} {\bibinfo {author} {\bibfnamefont {N.}~\bibnamefont
  {Andreou}}, \bibinfo {author} {\bibfnamefont {N.}~\bibnamefont {Franchini}},
  \bibinfo {author} {\bibfnamefont {G.}~\bibnamefont {Ventagli}},\ and\
  \bibinfo {author} {\bibfnamefont {T.~P.}\ \bibnamefont {Sotiriou}},\
  }\bibfield  {title} {\bibinfo {title} {{Spontaneous scalarization in
  generalized scalar-tensor theory}},\ }\href
  {https://doi.org/10.1103/PhysRevD.99.124022} {\bibfield  {journal} {\bibinfo
  {journal} {Phys. Rev.}\ }\textbf {\bibinfo {volume} {D99}},\ \bibinfo {pages}
  {124022} (\bibinfo {year} {2019})},\ \Eprint
  {https://arxiv.org/abs/1904.06365} {arXiv:1904.06365 [gr-qc]} \BibitemShut
  {NoStop}%
\bibitem [{\citenamefont {Antoniou}\ \emph
  {et~al.}(2021{\natexlab{a}})\citenamefont {Antoniou}, \citenamefont
  {Bordin},\ and\ \citenamefont {Sotiriou}}]{Antoniou:2020nax}%
  \BibitemOpen
  \bibfield  {author} {\bibinfo {author} {\bibfnamefont {G.}~\bibnamefont
  {Antoniou}}, \bibinfo {author} {\bibfnamefont {L.}~\bibnamefont {Bordin}},\
  and\ \bibinfo {author} {\bibfnamefont {T.~P.}\ \bibnamefont {Sotiriou}},\
  }\bibfield  {title} {\bibinfo {title} {{Compact object scalarization with
  general relativity as a cosmic attractor}},\ }\href
  {https://doi.org/10.1103/PhysRevD.103.024012} {\bibfield  {journal} {\bibinfo
   {journal} {Phys. Rev. D}\ }\textbf {\bibinfo {volume} {103}},\ \bibinfo
  {pages} {024012} (\bibinfo {year} {2021}{\natexlab{a}})},\ \Eprint
  {https://arxiv.org/abs/2004.14985} {arXiv:2004.14985 [gr-qc]} \BibitemShut
  {NoStop}%
\bibitem [{\citenamefont {Ventagli}\ \emph {et~al.}(2020)\citenamefont
  {Ventagli}, \citenamefont {Leh\'ebel},\ and\ \citenamefont
  {Sotiriou}}]{Ventagli:2020rnx}%
  \BibitemOpen
  \bibfield  {author} {\bibinfo {author} {\bibfnamefont {G.}~\bibnamefont
  {Ventagli}}, \bibinfo {author} {\bibfnamefont {A.}~\bibnamefont
  {Leh\'ebel}},\ and\ \bibinfo {author} {\bibfnamefont {T.~P.}\ \bibnamefont
  {Sotiriou}},\ }\bibfield  {title} {\bibinfo {title} {{Onset of spontaneous
  scalarization in generalized scalar-tensor theories}},\ }\href
  {https://doi.org/10.1103/PhysRevD.102.024050} {\bibfield  {journal} {\bibinfo
   {journal} {Phys. Rev. D}\ }\textbf {\bibinfo {volume} {102}},\ \bibinfo
  {pages} {024050} (\bibinfo {year} {2020})},\ \Eprint
  {https://arxiv.org/abs/2006.01153} {arXiv:2006.01153 [gr-qc]} \BibitemShut
  {NoStop}%
\bibitem [{\citenamefont {Antoniou}\ \emph
  {et~al.}(2021{\natexlab{b}})\citenamefont {Antoniou}, \citenamefont
  {Leh\'ebel}, \citenamefont {Ventagli},\ and\ \citenamefont
  {Sotiriou}}]{Antoniou:2021zoy}%
  \BibitemOpen
  \bibfield  {author} {\bibinfo {author} {\bibfnamefont {G.}~\bibnamefont
  {Antoniou}}, \bibinfo {author} {\bibfnamefont {A.}~\bibnamefont {Leh\'ebel}},
  \bibinfo {author} {\bibfnamefont {G.}~\bibnamefont {Ventagli}},\ and\
  \bibinfo {author} {\bibfnamefont {T.~P.}\ \bibnamefont {Sotiriou}},\
  }\bibfield  {title} {\bibinfo {title} {{Black hole scalarization with
  Gauss-Bonnet and Ricci scalar couplings}},\ }\href@noop {} {\  (\bibinfo
  {year} {2021}{\natexlab{b}})},\ \Eprint {https://arxiv.org/abs/2105.04479}
  {arXiv:2105.04479 [gr-qc]} \BibitemShut {NoStop}%
\bibitem [{\citenamefont {Shao}\ \emph {et~al.}(2017)\citenamefont {Shao},
  \citenamefont {Sennett}, \citenamefont {Buonanno}, \citenamefont {Kramer},\
  and\ \citenamefont {Wex}}]{Shao:2017gwu}%
  \BibitemOpen
  \bibfield  {author} {\bibinfo {author} {\bibfnamefont {L.}~\bibnamefont
  {Shao}}, \bibinfo {author} {\bibfnamefont {N.}~\bibnamefont {Sennett}},
  \bibinfo {author} {\bibfnamefont {A.}~\bibnamefont {Buonanno}}, \bibinfo
  {author} {\bibfnamefont {M.}~\bibnamefont {Kramer}},\ and\ \bibinfo {author}
  {\bibfnamefont {N.}~\bibnamefont {Wex}},\ }\bibfield  {title} {\bibinfo
  {title} {{Constraining nonperturbative strong-field effects in scalar-tensor
  gravity by combining pulsar timing and laser-interferometer
  gravitational-wave detectors}},\ }\href
  {https://doi.org/10.1103/PhysRevX.7.041025} {\bibfield  {journal} {\bibinfo
  {journal} {Phys. Rev. X}\ }\textbf {\bibinfo {volume} {7}},\ \bibinfo {pages}
  {041025} (\bibinfo {year} {2017})},\ \Eprint
  {https://arxiv.org/abs/1704.07561} {arXiv:1704.07561 [gr-qc]} \BibitemShut
  {NoStop}%
\bibitem [{\citenamefont {Wex}\ and\ \citenamefont
  {Kramer}(2020)}]{Wex:2020ald}%
  \BibitemOpen
  \bibfield  {author} {\bibinfo {author} {\bibfnamefont {N.}~\bibnamefont
  {Wex}}\ and\ \bibinfo {author} {\bibfnamefont {M.}~\bibnamefont {Kramer}},\
  }\bibfield  {title} {\bibinfo {title} {{Gravity Tests with Radio Pulsars}},\
  }\href {https://doi.org/10.3390/universe6090156} {\bibfield  {journal}
  {\bibinfo  {journal} {Universe}\ }\textbf {\bibinfo {volume} {6}},\ \bibinfo
  {pages} {156} (\bibinfo {year} {2020})}\BibitemShut {NoStop}%
\bibitem [{\citenamefont {Damour}\ and\ \citenamefont
  {Esposito-Farese}(1992)}]{Damour:1992we}%
  \BibitemOpen
  \bibfield  {author} {\bibinfo {author} {\bibfnamefont {T.}~\bibnamefont
  {Damour}}\ and\ \bibinfo {author} {\bibfnamefont {G.}~\bibnamefont
  {Esposito-Farese}},\ }\bibfield  {title} {\bibinfo {title} {{Tensor
  multiscalar theories of gravitation}},\ }\href
  {https://doi.org/10.1088/0264-9381/9/9/015} {\bibfield  {journal} {\bibinfo
  {journal} {Class. Quant. Grav.}\ }\textbf {\bibinfo {volume} {9}},\ \bibinfo
  {pages} {2093} (\bibinfo {year} {1992})}\BibitemShut {NoStop}%
\bibitem [{\citenamefont {Haensel}\ and\ \citenamefont
  {Potekhin}(2004)}]{Haensel:2004nu}%
  \BibitemOpen
  \bibfield  {author} {\bibinfo {author} {\bibfnamefont {P.}~\bibnamefont
  {Haensel}}\ and\ \bibinfo {author} {\bibfnamefont {A.~Y.}\ \bibnamefont
  {Potekhin}},\ }\bibfield  {title} {\bibinfo {title} {{Analytical
  representations of unified equations of state of neutron-star matter}},\
  }\href {https://doi.org/10.1051/0004-6361:20041722} {\bibfield  {journal}
  {\bibinfo  {journal} {Astron. Astrophys.}\ }\textbf {\bibinfo {volume}
  {428}},\ \bibinfo {pages} {191} (\bibinfo {year} {2004})},\ \Eprint
  {https://arxiv.org/abs/astro-ph/0408324} {arXiv:astro-ph/0408324}
  \BibitemShut {NoStop}%
\bibitem [{\citenamefont {Silva}\ \emph {et~al.}(2019)\citenamefont {Silva},
  \citenamefont {Macedo}, \citenamefont {Sotiriou}, \citenamefont {Gualtieri},
  \citenamefont {Sakstein},\ and\ \citenamefont {Berti}}]{Silva:2018qhn}%
  \BibitemOpen
  \bibfield  {author} {\bibinfo {author} {\bibfnamefont {H.~O.}\ \bibnamefont
  {Silva}}, \bibinfo {author} {\bibfnamefont {C.~F.}\ \bibnamefont {Macedo}},
  \bibinfo {author} {\bibfnamefont {T.~P.}\ \bibnamefont {Sotiriou}}, \bibinfo
  {author} {\bibfnamefont {L.}~\bibnamefont {Gualtieri}}, \bibinfo {author}
  {\bibfnamefont {J.}~\bibnamefont {Sakstein}},\ and\ \bibinfo {author}
  {\bibfnamefont {E.}~\bibnamefont {Berti}},\ }\bibfield  {title} {\bibinfo
  {title} {{Stability of scalarized black hole solutions in scalar-Gauss-Bonnet
  gravity}},\ }\href {https://doi.org/10.1103/PhysRevD.99.064011} {\bibfield
  {journal} {\bibinfo  {journal} {Phys. Rev. D}\ }\textbf {\bibinfo {volume}
  {99}},\ \bibinfo {pages} {064011} (\bibinfo {year} {2019})},\ \Eprint
  {https://arxiv.org/abs/1812.05590} {arXiv:1812.05590 [gr-qc]} \BibitemShut
  {NoStop}%
\bibitem [{\citenamefont {Gungor}\ and\ \citenamefont
  {Eksi}(2011)}]{Gungor:2011vq}%
  \BibitemOpen
  \bibfield  {author} {\bibinfo {author} {\bibfnamefont {C.}~\bibnamefont
  {Gungor}}\ and\ \bibinfo {author} {\bibfnamefont {K.~Y.}\ \bibnamefont
  {Eksi}},\ }\bibfield  {title} {\bibinfo {title} {{Analytical Representation
  for Equations of State of Dense Matter}},\ }\href@noop {} {\  (\bibinfo
  {year} {2011})},\ \Eprint {https://arxiv.org/abs/1108.2166} {arXiv:1108.2166
  [astro-ph.SR]} \BibitemShut {NoStop}%
\bibitem [{\citenamefont {Mendes}(2015)}]{Mendes:2014ufa}%
  \BibitemOpen
  \bibfield  {author} {\bibinfo {author} {\bibfnamefont {R.~F.~P.}\
  \bibnamefont {Mendes}},\ }\bibfield  {title} {\bibinfo {title} {{Possibility
  of setting a new constraint to scalar-tensor theories}},\ }\href
  {https://doi.org/10.1103/PhysRevD.91.064024} {\bibfield  {journal} {\bibinfo
  {journal} {Phys. Rev. D}\ }\textbf {\bibinfo {volume} {91}},\ \bibinfo
  {pages} {064024} (\bibinfo {year} {2015})},\ \Eprint
  {https://arxiv.org/abs/1412.6789} {arXiv:1412.6789 [gr-qc]} \BibitemShut
  {NoStop}%
\bibitem [{\citenamefont {Palenzuela}\ and\ \citenamefont
  {Liebling}(2016)}]{Palenzuela:2015ima}%
  \BibitemOpen
  \bibfield  {author} {\bibinfo {author} {\bibfnamefont {C.}~\bibnamefont
  {Palenzuela}}\ and\ \bibinfo {author} {\bibfnamefont {S.~L.}\ \bibnamefont
  {Liebling}},\ }\bibfield  {title} {\bibinfo {title} {{Constraining
  scalar-tensor theories of gravity from the most massive neutron stars}},\
  }\href {https://doi.org/10.1103/PhysRevD.93.044009} {\bibfield  {journal}
  {\bibinfo  {journal} {Phys. Rev. D}\ }\textbf {\bibinfo {volume} {93}},\
  \bibinfo {pages} {044009} (\bibinfo {year} {2016})},\ \Eprint
  {https://arxiv.org/abs/1510.03471} {arXiv:1510.03471 [gr-qc]} \BibitemShut
  {NoStop}%
\bibitem [{\citenamefont {Mendes}\ and\ \citenamefont
  {Ortiz}(2016)}]{Mendes:2016fby}%
  \BibitemOpen
  \bibfield  {author} {\bibinfo {author} {\bibfnamefont {R.~F.~P.}\
  \bibnamefont {Mendes}}\ and\ \bibinfo {author} {\bibfnamefont
  {N.}~\bibnamefont {Ortiz}},\ }\bibfield  {title} {\bibinfo {title} {{Highly
  compact neutron stars in scalar-tensor theories of gravity: Spontaneous
  scalarization versus gravitational collapse}},\ }\href
  {https://doi.org/10.1103/PhysRevD.93.124035} {\bibfield  {journal} {\bibinfo
  {journal} {Phys. Rev. D}\ }\textbf {\bibinfo {volume} {93}},\ \bibinfo
  {pages} {124035} (\bibinfo {year} {2016})},\ \Eprint
  {https://arxiv.org/abs/1604.04175} {arXiv:1604.04175 [gr-qc]} \BibitemShut
  {NoStop}%
\bibitem [{\citenamefont {Freire}\ \emph {et~al.}(2012)\citenamefont {Freire},
  \citenamefont {Wex}, \citenamefont {Esposito-Farese}, \citenamefont
  {Verbiest}, \citenamefont {Bailes}, \citenamefont {Jacoby}, \citenamefont
  {Kramer}, \citenamefont {Stairs}, \citenamefont {Antoniadis},\ and\
  \citenamefont {Janssen}}]{Freire:2012mg}%
  \BibitemOpen
  \bibfield  {author} {\bibinfo {author} {\bibfnamefont {P.~C.}\ \bibnamefont
  {Freire}}, \bibinfo {author} {\bibfnamefont {N.}~\bibnamefont {Wex}},
  \bibinfo {author} {\bibfnamefont {G.}~\bibnamefont {Esposito-Farese}},
  \bibinfo {author} {\bibfnamefont {J.~P.}\ \bibnamefont {Verbiest}}, \bibinfo
  {author} {\bibfnamefont {M.}~\bibnamefont {Bailes}}, \bibinfo {author}
  {\bibfnamefont {B.~A.}\ \bibnamefont {Jacoby}}, \bibinfo {author}
  {\bibfnamefont {M.}~\bibnamefont {Kramer}}, \bibinfo {author} {\bibfnamefont
  {I.~H.}\ \bibnamefont {Stairs}}, \bibinfo {author} {\bibfnamefont
  {J.}~\bibnamefont {Antoniadis}},\ and\ \bibinfo {author} {\bibfnamefont
  {G.~H.}\ \bibnamefont {Janssen}},\ }\bibfield  {title} {\bibinfo {title}
  {{The relativistic pulsar-white dwarf binary PSR J1738+0333 II. The most
  stringent test of scalar-tensor gravity}},\ }\href
  {https://doi.org/10.1111/j.1365-2966.2012.21253.x} {\bibfield  {journal}
  {\bibinfo  {journal} {Mon. Not. Roy. Astron. Soc.}\ }\textbf {\bibinfo
  {volume} {423}},\ \bibinfo {pages} {3328} (\bibinfo {year} {2012})},\ \Eprint
  {https://arxiv.org/abs/1205.1450} {arXiv:1205.1450 [astro-ph.GA]}
  \BibitemShut {NoStop}%
\bibitem [{\citenamefont {Antoniadis}\ \emph {et~al.}(2013)\citenamefont
  {Antoniadis} \emph {et~al.}}]{Antoniadis:2013pzd}%
  \BibitemOpen
  \bibfield  {author} {\bibinfo {author} {\bibfnamefont {J.}~\bibnamefont
  {Antoniadis}} \emph {et~al.},\ }\bibfield  {title} {\bibinfo {title} {{A
  Massive Pulsar in a Compact Relativistic Binary}},\ }\href
  {https://doi.org/10.1126/science.1233232} {\bibfield  {journal} {\bibinfo
  {journal} {Science}\ }\textbf {\bibinfo {volume} {340}},\ \bibinfo {pages}
  {6131} (\bibinfo {year} {2013})},\ \Eprint {https://arxiv.org/abs/1304.6875}
  {arXiv:1304.6875 [astro-ph.HE]} \BibitemShut {NoStop}%
\bibitem [{\citenamefont {Maselli}\ \emph {et~al.}(2020)\citenamefont
  {Maselli}, \citenamefont {Franchini}, \citenamefont {Gualtieri},\ and\
  \citenamefont {Sotiriou}}]{Maselli:2020zgv}%
  \BibitemOpen
  \bibfield  {author} {\bibinfo {author} {\bibfnamefont {A.}~\bibnamefont
  {Maselli}}, \bibinfo {author} {\bibfnamefont {N.}~\bibnamefont {Franchini}},
  \bibinfo {author} {\bibfnamefont {L.}~\bibnamefont {Gualtieri}},\ and\
  \bibinfo {author} {\bibfnamefont {T.~P.}\ \bibnamefont {Sotiriou}},\
  }\bibfield  {title} {\bibinfo {title} {{Detecting scalar fields with Extreme
  Mass Ratio Inspirals}},\ }\href
  {https://doi.org/10.1103/PhysRevLett.125.141101} {\bibfield  {journal}
  {\bibinfo  {journal} {Phys. Rev. Lett.}\ }\textbf {\bibinfo {volume} {125}},\
  \bibinfo {pages} {141101} (\bibinfo {year} {2020})},\ \Eprint
  {https://arxiv.org/abs/2004.11895} {arXiv:2004.11895 [gr-qc]} \BibitemShut
  {NoStop}%
\bibitem [{\citenamefont {Maselli}\ \emph {et~al.}(2021)\citenamefont
  {Maselli}, \citenamefont {Franchini}, \citenamefont {Gualtieri},
  \citenamefont {Sotiriou}, \citenamefont {Barsanti},\ and\ \citenamefont
  {Pani}}]{Maselli:2021men}%
  \BibitemOpen
  \bibfield  {author} {\bibinfo {author} {\bibfnamefont {A.}~\bibnamefont
  {Maselli}}, \bibinfo {author} {\bibfnamefont {N.}~\bibnamefont {Franchini}},
  \bibinfo {author} {\bibfnamefont {L.}~\bibnamefont {Gualtieri}}, \bibinfo
  {author} {\bibfnamefont {T.~P.}\ \bibnamefont {Sotiriou}}, \bibinfo {author}
  {\bibfnamefont {S.}~\bibnamefont {Barsanti}},\ and\ \bibinfo {author}
  {\bibfnamefont {P.}~\bibnamefont {Pani}},\ }\bibfield  {title} {\bibinfo
  {title} {{Detecting new fundamental fields with LISA}},\ }\href@noop {} {\
  (\bibinfo {year} {2021})},\ \Eprint {https://arxiv.org/abs/2106.11325}
  {arXiv:2106.11325 [gr-qc]} \BibitemShut {NoStop}%
\bibitem [{\citenamefont {Ripley}\ and\ \citenamefont
  {Pretorius}(2020)}]{Ripley:2020vpk}%
  \BibitemOpen
  \bibfield  {author} {\bibinfo {author} {\bibfnamefont {J.~L.}\ \bibnamefont
  {Ripley}}\ and\ \bibinfo {author} {\bibfnamefont {F.}~\bibnamefont
  {Pretorius}},\ }\bibfield  {title} {\bibinfo {title} {{Dynamics of a $\mathbb
  Z_2$ symmetric EdGB gravity in spherical symmetry}},\ }\href
  {https://doi.org/10.1088/1361-6382/ab9bbb} {\bibfield  {journal} {\bibinfo
  {journal} {Class. Quant. Grav.}\ }\textbf {\bibinfo {volume} {37}},\ \bibinfo
  {pages} {155003} (\bibinfo {year} {2020})},\ \Eprint
  {https://arxiv.org/abs/2005.05417} {arXiv:2005.05417 [gr-qc]} \BibitemShut
  {NoStop}%
\bibitem [{\citenamefont {Cunha}\ \emph {et~al.}(2019)\citenamefont {Cunha},
  \citenamefont {Herdeiro},\ and\ \citenamefont {Radu}}]{Cunha:2019dwb}%
  \BibitemOpen
  \bibfield  {author} {\bibinfo {author} {\bibfnamefont {P.~V.~P.}\
  \bibnamefont {Cunha}}, \bibinfo {author} {\bibfnamefont {C.~A.~R.}\
  \bibnamefont {Herdeiro}},\ and\ \bibinfo {author} {\bibfnamefont
  {E.}~\bibnamefont {Radu}},\ }\bibfield  {title} {\bibinfo {title}
  {{Spontaneously Scalarized Kerr Black Holes in Extended
  Scalar-Tensor\textendash{}Gauss-Bonnet Gravity}},\ }\href
  {https://doi.org/10.1103/PhysRevLett.123.011101} {\bibfield  {journal}
  {\bibinfo  {journal} {Phys. Rev. Lett.}\ }\textbf {\bibinfo {volume} {123}},\
  \bibinfo {pages} {011101} (\bibinfo {year} {2019})},\ \Eprint
  {https://arxiv.org/abs/1904.09997} {arXiv:1904.09997 [gr-qc]} \BibitemShut
  {NoStop}%
\bibitem [{\citenamefont {Collodel}\ \emph {et~al.}(2020)\citenamefont
  {Collodel}, \citenamefont {Kleihaus}, \citenamefont {Kunz},\ and\
  \citenamefont {Berti}}]{Collodel:2019kkx}%
  \BibitemOpen
  \bibfield  {author} {\bibinfo {author} {\bibfnamefont {L.~G.}\ \bibnamefont
  {Collodel}}, \bibinfo {author} {\bibfnamefont {B.}~\bibnamefont {Kleihaus}},
  \bibinfo {author} {\bibfnamefont {J.}~\bibnamefont {Kunz}},\ and\ \bibinfo
  {author} {\bibfnamefont {E.}~\bibnamefont {Berti}},\ }\bibfield  {title}
  {\bibinfo {title} {{Spinning and excited black holes in
  Einstein-scalar-Gauss\textendash{}Bonnet theory}},\ }\href
  {https://doi.org/10.1088/1361-6382/ab74f9} {\bibfield  {journal} {\bibinfo
  {journal} {Class. Quant. Grav.}\ }\textbf {\bibinfo {volume} {37}},\ \bibinfo
  {pages} {075018} (\bibinfo {year} {2020})},\ \Eprint
  {https://arxiv.org/abs/1912.05382} {arXiv:1912.05382 [gr-qc]} \BibitemShut
  {NoStop}%
\bibitem [{\citenamefont {Doneva}\ \emph {et~al.}(2013)\citenamefont {Doneva},
  \citenamefont {Yazadjiev}, \citenamefont {Stergioulas},\ and\ \citenamefont
  {Kokkotas}}]{Doneva:2013qva}%
  \BibitemOpen
  \bibfield  {author} {\bibinfo {author} {\bibfnamefont {D.~D.}\ \bibnamefont
  {Doneva}}, \bibinfo {author} {\bibfnamefont {S.~S.}\ \bibnamefont
  {Yazadjiev}}, \bibinfo {author} {\bibfnamefont {N.}~\bibnamefont
  {Stergioulas}},\ and\ \bibinfo {author} {\bibfnamefont {K.~D.}\ \bibnamefont
  {Kokkotas}},\ }\bibfield  {title} {\bibinfo {title} {{Rapidly rotating
  neutron stars in scalar-tensor theories of gravity}},\ }\href
  {https://doi.org/10.1103/PhysRevD.88.084060} {\bibfield  {journal} {\bibinfo
  {journal} {Phys. Rev. D}\ }\textbf {\bibinfo {volume} {88}},\ \bibinfo
  {pages} {084060} (\bibinfo {year} {2013})},\ \Eprint
  {https://arxiv.org/abs/1309.0605} {arXiv:1309.0605 [gr-qc]} \BibitemShut
  {NoStop}%
\end{thebibliography}%

\end{document}